\documentclass[english,journal=ancac3,manuscript=article,layout=twocolumn]{achemso}
\usepackage[T1]{fontenc}
\usepackage[latin9]{inputenc}
\usepackage{amsmath}
\usepackage{amssymb}
\usepackage{graphicx}

\let\oldmaketitle\maketitle
\let\maketitle\relax

\makeatletter


\title{Circumventing the Phonon Bottleneck by Multiphonon-Mediated Hot Exciton Cooling at the Nanoscale}

\author{Dipti Jasrasaria}
\affiliation{Department of Chemistry, University of California, Berkeley, California 94720, United States}
\altaffiliation{Current address: Department of Chemistry, Columbia University, New York, New York 10027, United States; dj2667@columbia.edu}
\email{djasrasaria@berkeley.edu}

\author{Eran Rabani}
\affiliation{Department of Chemistry, University of California, Berkeley, California 94720, United States}
\alsoaffiliation{Materials Sciences Division, Lawrence Berkeley National Laboratory, Berkeley, California 94720, United States}
\alsoaffiliation{The Sackler Center for Computational Molecular and Materials Science, Tel Aviv University, Tel Aviv, Israel 69978}
\email{eran.rabani@berkeley.edu}


\setkeys{acs}{articletitle = true}
\usepackage{bm}
\usepackage{xcolor}
\usepackage{soul}

\makeatother

\usepackage{babel}
\begin{document}

\twocolumn[
\begin{@twocolumnfalse}
\oldmaketitle
\begin{abstract}
In semiconductor materials, hot exciton cooling is the process by which highly excited carriers nonradiatively relax to form a band edge exciton. While cooling plays an important role in determining the thermal losses and quantum yield of a system, the timescales and mechanism of cooling are not well understood in confined semiconductor nanocrystals (NCs). A mismatch between electronic energy gaps and phonon frequencies in NCs has led to the hypothesis of a phonon bottleneck, in which cooling would be extremely slow, while enhanced electron-hole interactions in NCs have been used to explain cooling that would occur on ultrafast timescales. Here, we develop an atomistic approach for describing phonon-mediated exciton dynamics, and we use it to simulate hot exciton cooling in NCs of experimentally relevant sizes. Our framework includes electron-hole correlations as well as multiphonon processes, both of which are necessary to accurately describe the cooling process. We find that cooling occurs on timescales of tens of femtoseconds in CdSe cores, in agreement with experimental measurements, through a cascade of relaxation events that are mediated by efficient multiphonon emission. Cooling timescales increase with increasing NC size due to decreased exciton-phonon coupling (EXPC), and they are an order of magnitude larger in CdSe-CdS core-shell NCs because of reduced EXPC to low- and mid-frequency acoustic modes.
\end{abstract}
\end{@twocolumnfalse}
]


Understanding the mechanisms of nonradiative decay of electronic excited states in semiconductor nanocrystals (NCs) is key to developing NC-based technologies with decreased thermal losses and increased device efficiencies.\cite{Lannoo1996, klimov2006mechanisms, sheik2007optical, fomenko2008solution, Talapin2010, Hanifi2019} When a NC is excited by a photon with an energy larger than that of the NC band gap, the absorbed photon generates a highly excited electron-hole pair. The process by which these excited, or ``hot'', carriers nonradiatively relax to form a band edge exciton is often referred to as~\cite{Weiss2014, PGS2021} ``hot exciton cooling.'' In bulk semiconductors, Fr\"{o}hlich and deformation potential interactions between excitons and phonons as well as continuous densities of electronic and phonon states allow for efficient hot exciton cooling that occurs on timescales of $\sim$1\,ps or less.\cite{Lambrich1979, Pugnet1981, Prabhu1995} However, in semiconductor NCs, confinement changes the nature of exciton-phonon coupling (EXPC) and leads to the discretization of both electronic and phonon states. These qualitative changes have led to open questions regarding the timescales and mechanisms of hot exciton cooling in confined semiconductor materials.\cite{Kambhampati2011, Knowles2011, Weiss2014, PGS2021}

\begin{figure}[h!]
\centering
\includegraphics{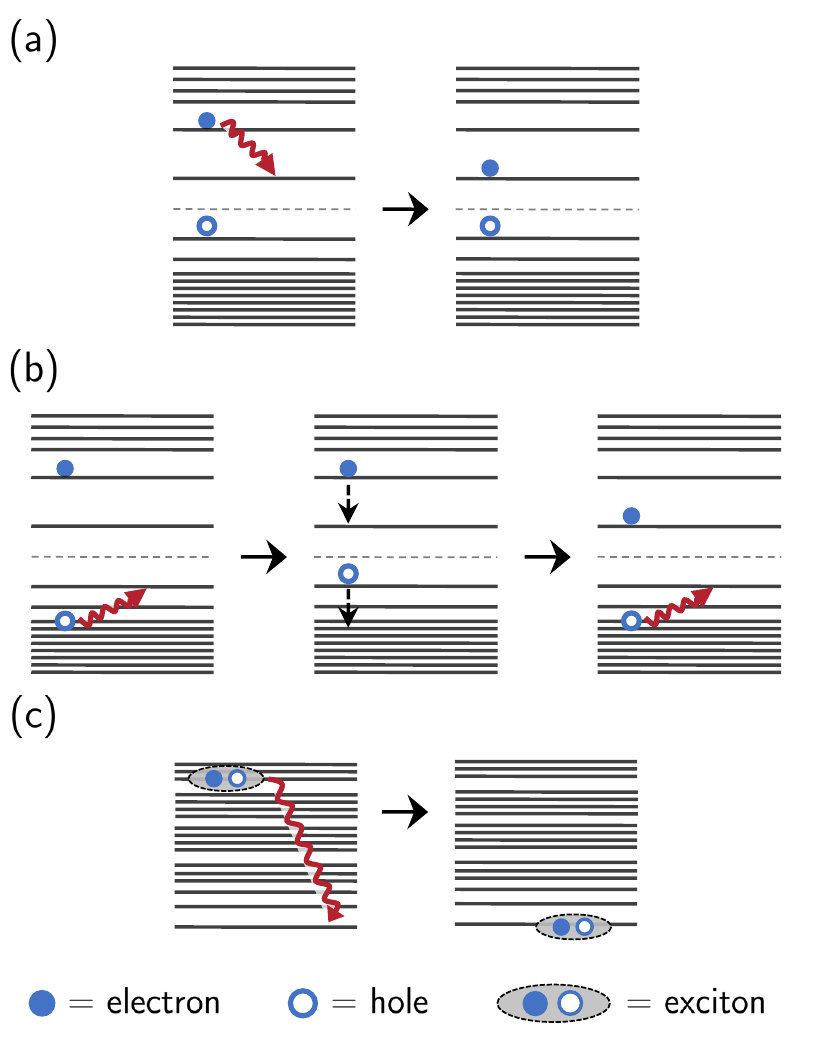}
\caption{(a) The phonon bottleneck refers to phonon-mediated, hot electron relaxation near the band edge, which would require the simultaneous emission of several of phonons due to the large energy gaps between these single-particle electronic states.
(b) The Auger-assisted cooling mechanism involves Coulomb-mediated interactions between electron and hole states. The hole first relaxes to the band edge \textit{via} phonon emission. Then, the electron relaxes by re-exciting the hole, which then relaxes to the band edge once more.
(c) Our formalism includes both electron-hole correlations and exciton-phonon interactions, circumventing the phonon bottleneck through a cascade of phonon-mediated transitions between \textit{excitonic} states with smaller energy gaps.}
\label{Fig1_Mechanism}
\end{figure}

In a picture of non-interacting electrons and holes, the hot electron and hot hole would relax independently from one another. The hole, which has a heavier effective mass than the electron in most II-VI and III-V semiconductors, has a higher density of states with energy gaps that are on the order of the phonon frequencies in the system.\cite{Jasrasaria2022_JCP} Thus, resonance conditions required for hole relaxation \textit{via} single phonon emission can be easily satisfied, and the hole can relax quickly to the band edge. The electron, however, has energy gaps, especially near the conduction band edge, that can be hundreds of meV, which is an order of magnitude larger than the typical optical phonon frequencies in the system.\cite{Jasrasaria2022_JCP} This energy mismatch has led to the hypothesis of a phonon bottleneck in NCs,\cite{Nozik2001} where hot electron cooling \textit{via} phonon emission would require a multiphonon process, as depicted in Fig.~\ref{Fig1_Mechanism}a. The simultaneous emission of tens of phonons would be inefficient,\cite{Inoshita1992} leading to very slow relaxation.

Experimental measurements of this cooling process in NCs rely on time-resolved spectroscopy, such as transient absorption, and have yielded conflicting results. Some experiments show slow relaxation that occurs on timescales of 10\,ps or longer,\cite{Gfroerer1996, Haiping1996, Heitz1997, GS1999, Sosnowski1998, Mukai1998, Pandey2008} which support the phonon bottleneck hypothesis. These measurements were primarily performed on larger, self-assembled III-V NCs,\cite{asahi1997self} which are in the weak confinement regime and which tend to have many localized trap states associated with structural defects. Other experiments, especially on colloidal II-VI NCs in the strong confinement regime, have observed relaxation that occurs within hundreds of femtoseconds with carrier energy loss rates that are much faster than those of bulk carriers.\cite{klimov1998femtosecond, Klimov1999, Klimov2000a, Schaller2005, Harbold2005, KambhampatiPRB2007, KambhampatiPRL2007}

This fast relaxation of strongly confined electron-hole pairs was attributed to an Auger-assisted cooling mechanism that circumvents the phonon bottleneck through Coulomb-mediated interactions between the electron and hole.\cite{Kharchenko1996, Efros1995} In this mechanism, illustrated schematically in Fig.~\ref{Fig1_Mechanism}b, the hot hole quickly relaxes to the band edge \textit{via} phonon emission. Then, the hot electron relaxes to the band edge by nonradiatively re-exciting the hole in an Auger-like process, and finally the hole relaxes again. The Auger cooling mechanism has been supported by observations that relaxation is faster in smaller NCs,\cite{klimov1998femtosecond, Klimov1999, Schaller2005, Hendry2006, KambhampatiPRB2007} in which electron-hole correlations are larger and Auger rates are faster.\cite{Philbin2018} Additionally, relaxation timescales increase drastically in NCs that are passivated with hole-accepting pyridine ligands,\cite{GS1999, Klimov2000a, GS2005} indicating that electron-hole correlations are important in the cooling mechanism. Other studies suggest that fast relaxation is due to efficient multiphonon emission\cite{Schaller2005} or due to coupling of electrons and/or holes to the vibrational modes of surface passivating ligands.\cite{GS2005, KambhampatiPRL2007}

While the Auger cooling mechanism provides an avenue for breaking the phonon bottleneck, it lacks essential physics for a complete description of hot exciton cooling in NCs. For example, electron-hole interactions, which are enhanced in confined semiconductors,\cite{Jasrasaria2022_JCP} are only considered perturbatively. Furthermore, the Auger cooling mechanism lacks the mechanistic details of the rapid hole relaxation and, for certain systems,~\cite{Klimov2002,KambhampatiPRL2007} may result in a hole phonon bottleneck when multiphonon relaxation pathways are assumed to be negligible. 

The computational challenges~\cite{Wang2003, Kilina2009, Oleg2009, Zeng2021} associated with accurately calculating excitons and their phonon-mediated dynamics in systems with thousands of valence electrons and hundreds of atoms have made it difficult to delineate the mechanism of hot exciton cooling and its dependence on NC properties, such as size and material composition.\cite{Weiss2014, PGS2021} A fundamental understanding of this process may offer rational design principles for NCs with tuned cooling timescales that are optimized for different NC-based applications.\cite{Kambhampati2011, fomenko2008solution, Pandey2008, pandey2010hot}

Here, we develop an atomistic theory to describe hot exciton cooling in II-VI NCs of experimentally relevant sizes. Our framework describes phonon-mediated transitions between \textit{excitonic} states, which inherently include electron-hole correlations (Fig.~\ref{Fig1_Mechanism}c). Furthermore, we accurately describe the exciton-phonon couplings (EXPC)~\cite{Jasrasaria2021,Jasrasaria2022erratum} and include multiphonon-mediated excitonic transitions. We use a master equation approach, which assumes weak EXPC, to propagate exciton population dynamics.

The timescales and exciton decay mechanism emerge naturally in our approach. We find that cooling occurs on timescales of tens of femtoseconds in wurtzite CdSe NCs, in agreement with measurements,\cite{Klimov1999,KambhampatiPRB2007} and occurs an order of magnitude slower in wurtzite CdSe-CdS core-shell NCs due to the weaker EXPC in these systems. We show that this ultrafast timescale is governed by \textit{both} electron-hole correlations \textit{and} multiphonon emission processes, which are made efficient by the large number and quasi-continuous phonon modes in NCs. Our results are consistent with the picture emerging from the Auger-assisted relaxation mechanism, but, in addition, we attribute the lack of a phonon bottleneck to the important role of multiphonon emission processes.

\section{Results and Discussion}

\subsection{Describing phonon-mediated exciton cooling}

We adopt the following Hamiltonian to describe a manifold of excitonic states coupled to vibrational modes, with EXPC expanded to lowest order in the atomic displacements:\cite{Jasrasaria2021}
\begin{align}
    H = &\sum_{n}E_{n}\left|\psi_{n}\right\rangle \left\langle \psi_{n}\right|+\sum_{\alpha}\hbar\omega_{\alpha}b_{\alpha}^{\dagger}b_{\alpha} \nonumber \\
    &+\sum_{\alpha nm}V_{n,m}^{\alpha}\left|\psi_{n}\right\rangle \left\langle \psi_{m}\right|q_{\alpha}\,.
    \label{eq:og_Hamiltonian}
\end{align}
The excitonic energies, $E_n$, and states, $\vert\psi_n\rangle$, as well as the EXPC matrix elements, $V_{n,m}^\alpha$, were calculated using the semiempirical pseudopotential method coupled with the Bethe-Salpeter equation (see Jasrasaria \textit{et al.}~\cite{Jasrasaria2021, Jasrasaria2022_JCP} and the SI for more details). Phonon modes and frequencies, $\omega_\alpha$, were obtained by diagonalizing the dynamical matrix computed using a previously-parameterized atomic force field.\cite{Zhou2013}

\begin{figure*}[h!]
\centering
\includegraphics{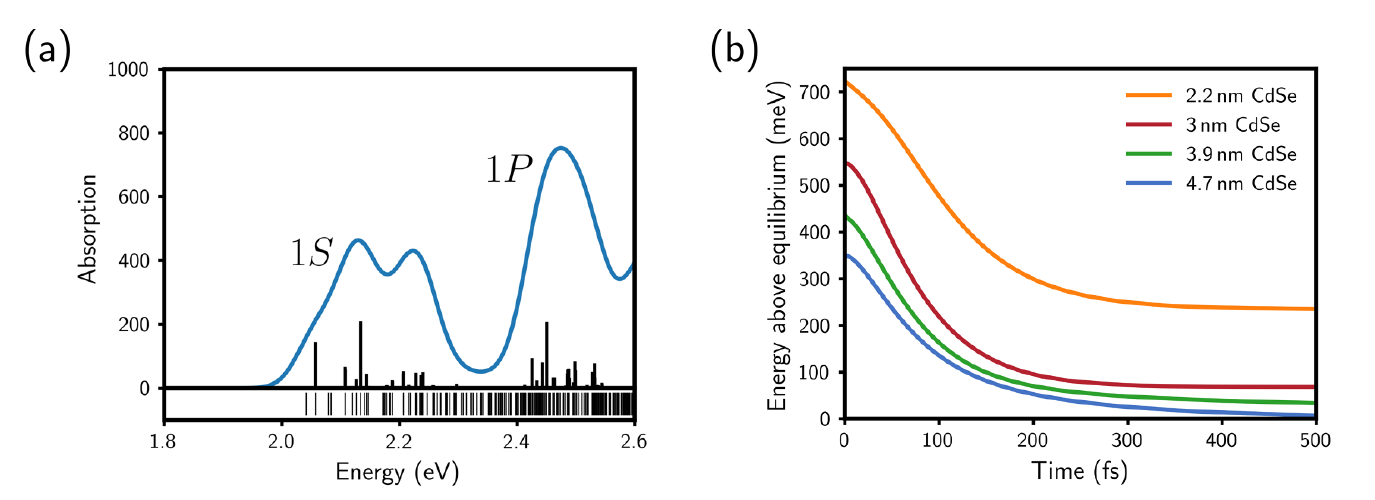}
\caption{(a) The calculated linear absorption spectrum (top) and density of excitonic states (bottom) for a 3.9\,nm CdSe NC. The vertical lines in the top panel indicate the magnitude of the oscillator strength of the transition from the ground state to that excitonic state. Cooling from the $1P$ to the $1S$ excitonic states, whose transitions are labeled in the absorption spectrum, is the main subject of this work.
(b) Single phonon-mediated hot exciton cooling simulated for CdSe NCs of different sizes shows a phonon bottleneck in smaller NCs because excitonic energy gaps in those systems are larger than the highest phonon frequencies.}
\label{Fig2_AbsorptionCooling}
\end{figure*}

In Fig.~\ref{Fig2_AbsorptionCooling}a we show the density of excitonic states scaled by the oscillator strengths for a typical CdSe NC with a diameter of 3.9\,nm as well as the corresponding linear absorption spectrum. We find that the underlying density of excitonic states is relatively high due to the dense spectrum of hole states. Some of these excitonic states correspond to \textit{bright} transitions from the ground state with large oscillator strengths while others correspond to \textit{dim} transitions for which the oscillator strengths are small. Note that we only show the bright/dim states, and the dark (forbidden) transitions are not shown. The linear absorption spectrum shows several distinct features, in agreement with experiments,\cite{Klimov1999, Sewall2006} that are governed by a few excitonic transitions with large oscillator strengths. We label the main transitions as $1S$ and $1P$, following the literature convention. With the dense manifold of excitonic states, the relaxation from the $1P$ excitonic state, in which the exciton electron is primarily composed of $p$-like, single-particle electron states, to the $1S$ ground excitonic state, in which both the exciton electron and hole are primarily comprised of band edge single-particle states, should occur through a cascade of phonon-mediated transitions, where both bright and dim excitonic states are involved.

We first consider the limit of single-phonon processes. We focused on the population dynamics and computed the phonon-mediated transition rates between excitonic states $n$ and $m$ using Fermi's golden rule, which assumes weak system-bath coupling and employs the Markovian approximation:
\begin{align}
    \Gamma_{n\rightarrow m} = &\frac{1}{\hbar^2} \int_{-\infty}^\infty dt e^{i(E_n - E_m)t/\hbar} \nonumber \\
    &\times \sum_\alpha V_{n,m}^\alpha V_{m,n}^\alpha \langle q_\alpha (t) q_\alpha (0) \rangle_\text{eq}\,,
    \label{eq:rates_singlePhonon}
\end{align}
where $\langle \dots \rangle_\text{eq}$ denotes an equilibrium average over bath coordinates. We computed these rates for all excitonic transitions and used them to build a kinetic master equation and propagate phonon-mediated exciton dynamics. We then calculated the average energy above equilibrium, $\langle \Delta E(t) \rangle = \sum_n E_n \big(p_n(t) - p_{n,\text{eq}} \big)$, where $p_n(t)$ is the population of state $n$ at time $t$ and $p_{n,\text{eq}}$ is the population of state $n$ at thermal equilibrium.

Because Eq.~(\ref{eq:og_Hamiltonian}) describes the EXPC to first order in the phonon mode coordinates, the Fermi's golden rule rates given by Eq.~(\ref{eq:rates_singlePhonon}) only account for excitonic transitions that occur \textit{via} the absorption or emission of a single phonon. While the largest energy gaps between excitonic states are an order of magnitude smaller than those between single-particle, electron states, they can still be larger than the phonon frequencies. Thus, transitions between those excitonic states would require the simultaneous emission of multiple phonons. Indeed, single-phonon-mediated cooling simulations for CdSe NCs of different sizes shows a phonon bottleneck (Fig.~\ref{Fig2_AbsorptionCooling}b). This phonon bottleneck is especially significant in smaller NCs for which confinement results in larger excitonic energy gaps, particularly at low excitonic energies, preventing the hot exciton from fully relaxing to the band edge through single-phonon emission.

\subsection{Multiphonon emission}

To account for multiphonon processes and maintain the simplicity of the master equation, we performed a unitary polaron transformation~\cite{nitzan2006chemical, SilbeyPolaronTransform2012, Xu2016, Franchini2021} to the Hamiltonian in Eq.~(\ref{eq:og_Hamiltonian}):
\begin{equation}
    \tilde{H} = e^S H e^{-S}\,,
\end{equation}
where
\begin{equation}
    S = - \frac{i}{\hbar} \sum_{\alpha k} \omega_\alpha^{-2} p_\alpha V_{k,k}^\alpha \vert \psi_k \rangle \langle \psi_k \vert
\end{equation}
and $p_\alpha$ is the momentum of phonon mode $\alpha$. A detailed derivation and description of the polaron-transformed Hamiltonian and its consequences are given in the Supporting Information.

With respect to the polaron-transformed Hamiltonian, Fermi's golden rule transition rates can be computed as
\begin{align}
    \Gamma_{n\rightarrow m}(t) = &\frac{1}{\hbar^2} \int_{0}^t d\tau e^{i(\varepsilon_n - \varepsilon_m)\tau/\hbar} \nonumber \\
    &\times \langle g_{n,m} (\tau) g_{m,n} (0) \rangle_\text{eq}\,,
    \label{eq:rates_multiPhonon}
\end{align}
where $\varepsilon_n \equiv E_n - \lambda_n$ is the energy of exciton $n$ scaled by its reorganization energy, and $g_{n,m} \equiv \sum_\alpha \tilde{V}_{n,m}^\alpha q_\alpha - \tilde{\lambda}_{nm}$ is the coupling between the polaronic states $n$ and $m$ (see the SI for more details). The Markovian approximation is not necessarily valid for this polaron-transformed Hamiltonian, as described in the Methods, so we compute the time-dependent, non-Markovian rates. Again, we computed all transition rates to build a kinetic master equation and propagate dynamics. Within this framework, we calculate the average energy above thermal equilibrium as before; namely, as $\langle \Delta \varepsilon(t) \rangle = \sum_n \varepsilon_n \big(p_n(t) - p_{n,\text{eq}} \big)$.

\begin{figure*}[h!]
\includegraphics{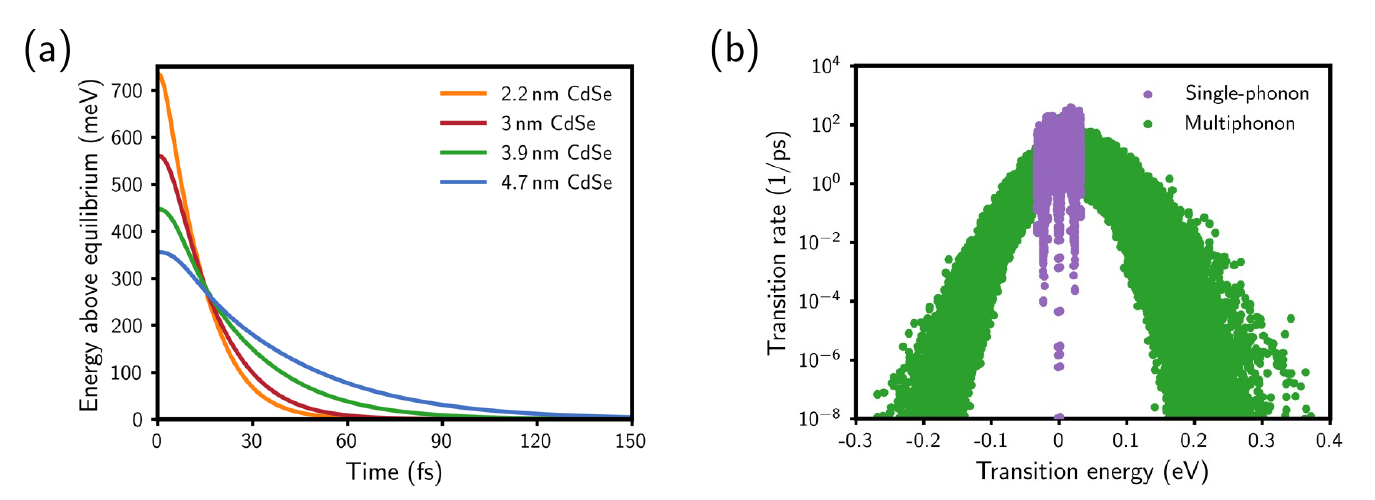}
\caption{(a) Hot exciton cooling simulated for CdSe cores of different sizes. The inclusion of multiphonon processes allows all systems to relax to thermal equilibrium, eliminating the phonon bottleneck.
(b) Exciton transition rates calculated for a 3.9\,nm CdSe NC within the single-phonon and multiphonon schemes. The single-phonon rates vanish for transition energies larger than the greatest phonon frequency while the multiphonon rates cover a wide range of transition energies.}
\label{Fig3_MultiphononCores}
\end{figure*}

Note that $g_{n,m}$ includes exponential functions of the phonon momenta, so multiphonon-mediated transitions are accounted for even in the lowest-order perturbation theory rate given by Eq.~(\ref{eq:rates_multiPhonon}). Including multiphonon processes enables transitions between excitonic states that have energy differences that are larger than the highest-frequency phonons. Multiphonon-mediated cooling simulations show that all NC systems fully relax to thermal equilibrium (Fig.~\ref{Fig3_MultiphononCores}a), indicating that multiphonon transitions involving a few phonon modes are essential to the cooling mechanism. Furthermore, the average energy relaxes within 100\,fs, much faster within the single-phonon scheme.

Examining the transition rate as a function of transition energy for a 3.9\,nm CdSe NC, illustrated in Fig.~\ref{Fig3_MultiphononCores}b, demonstrates that the single-phonon rates are larger for low-energy transitions, but there are no single-phonon transitions between excitonic states that have energy differences greater than $\sim$32\,meV (\textit{i.e.}, greater than the largest phonon energy). The multiphonon rates, however, cover the full range of transition energies. Importantly, multiphonon relaxation between excitonic states with energy differences of 100\,meV or less consistently have relatively high rates (ranging from $10^{-3}-10^2$ ps$^{-1}$ for multiphonon transitions as compared to $0-10^2$ ps$^{-1}$ for single-phonon transitions). This difference makes accessible many more relaxation channels and leads to a cooling timescale that is an order of magnitude faster than that resulting from single-phonon processes alone. The fast multiphonon relaxation is a result of the large number of phonon modes (approximately 3000 modes for a 3.9\,nm CdSe NC) that quasi-continuously span a wide frequency range and that are all coupled, to some degree, to excitonic transitions. Thus, many phonon combinations satisfy the energy conservation requirement for phonon-mediated exciton transitions, leading to efficient relaxation \textit{via} the emission of multiple phonons. 

The asymmetry in rates about 0\,meV transition energy reflects detailed balance (see Methods for more details). Furthermore, Fig.~\ref{Fig3_MultiphononCores}b shows a Gaussian relationship between the transition energy and the transition rate instead of the exponential dependence of the rate on the energy that results from the assumption that only the highest-frequency modes participate in nonradiative transitions.\cite{EnglmanJortner1970} This result indicates that lower-frequency acoustic \textit{and} optical modes are important to this cooling process, extending previous expectations that only high-frequency optical modes would be responsible for cooling.\cite{Nozik2001, Schaller2005}

Fig.~\ref{Fig3_MultiphononCores}a also demonstrates that smaller NCs relax more quickly to thermal equilibrium than larger NCs. Due to stronger quantum confinement, smaller NCs have more energy to dissipate during the cooling process. Smaller NCs also have larger excitonic gaps and a smaller number of phonon modes. However, smaller NCs have stronger EXPC than larger NCs,\cite{Jasrasaria2021} resulting in overall faster cooling timescales for smaller NCs.

\subsection{Controlling the cooling timescales}

\begin{figure*}[h!]
\includegraphics{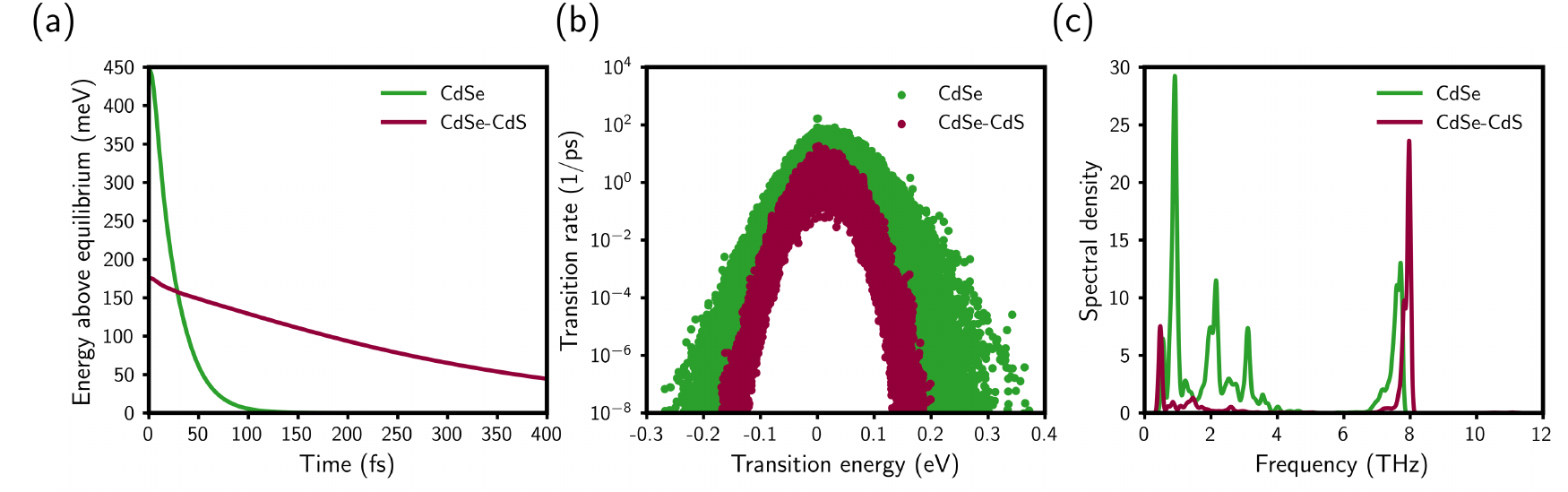}
\caption{(a) Hot exciton cooling simulated for a 3.9\,nm CdSe NC and a 3.9\,nm~CdSe--3\,ML~CdS core-shell NC. Cooling in the core-shell NC occurs five times slower due to reduced EXPC.
(b) Multiphonon transition rates calculated for the same CdSe and CdSe-CdS core-shell NCs, indicating that the rates for the core-shell NC are an order of magnitude smaller than that of the bare CdSe NC.
(c) Calculated spectral densities for the same CdSe and CdSe-CdS core-shell NCs. The core-shell NC shows stronger EXPC to the CdSe optical modes (around 7.5\,THz) than the bare CdSe NC. However, the bare CdSe NC has stronger coupling to lower- and mid-frequency modes, demonstrating that exciton coupling to a quasi-continuous frequency range of phonons is essential for efficient multiphonon relaxation.}
\label{Fig4_CoreShell}
\end{figure*}

The longer cooling timescales for larger CdSe NCs suggests that controlling the magnitude of EXPC may allow for tuning of the hot exciton cooling timescale. CdSe-CdS core-shell NCs have EXPC that is about five times smaller than that of bare cores due to suppression of exciton coupling to lower-frequency surface modes.\cite{Jasrasaria2021} Fig.~\ref{Fig4_CoreShell}a compares simulations of the cooling process for a 3.9\,nm CdSe core and a 3.9\,nm CdSe core with 3 monolayers of CdS shell. Because of the quasi-type II band alignment in CdSe-CdS core-shell NCs, the exciton hole is confined to the CdSe core while the electron somewhat delocalizes into the CdS shell.\cite{Eshet2013} This decreased quantum confinement leads to a smaller $1P$-$1S$ excitonic gap in core-shell NCs, so hot excitons have less energy to dissipate in core-shell NCs than in bare cores. However, cooling still takes about five times longer in the core-shell NC as a result of the weaker EXPC.

The multiphonon transition rates for both systems are shown in Fig.~\ref{Fig4_CoreShell}b, demonstrating that rates are one or more orders of magnitude smaller in the core-shell NC than in the bare core. For transitions with energies of 100\,meV for less, the multiphonon relaxation rates range from $10^{-3}-10^{2}$ ps$^{-1}$ for the CdSe NC and from $10^{-2}-10^1$ ps$^{-1}$ for the CdSe-CdS core-shell NC.

We can further understand the role of EXPC in the cooling mechanism by examining the spectral densities, which describe the phonon densities of states weighted by the EXPC, of the core and core-shell NCs (Fig.~\ref{Fig4_CoreShell}c). The spectral density of the CdSe core shows significant EXPC to acoustic modes with frequencies of 4.0\,THz or less as well as optical modes between 7.5$-$8.0\,THz.\cite{Jasrasaria2021,Jasrasaria2022erratum} The core-shell NC, however, has negligible EXPC at lower phonon frequencies, as the presence of the CdS shell prevents coupling to delocalized and surface-localized modes at those frequencies, but it has slightly stronger coupling to the CdSe optical modes. These results provide further evidence that both acoustic and optical modes play an essential role in the cooling process. Exciton coupling to phonons with a quasi-continuous frequency range allows for resonance conditions to be more easily satisfied; for every excitonic energy gap, a set of phonons with the corresponding energy is easily found. However, exciton coupling to phonons within a narrow energy range restrict the set of phonons that would satisfy the necessary resonance conditions.

\subsection{Energy loss rates}

To allow for more meaningful comparison between our calculations and experimental measurements, we simulate changes in the absorption spectrum of a system initially excited to the $1P$ excitonic state as it relaxes to the $1S$ ground excitonic state.
Assuming that the electric field, $\mathcal{E}$, is weak, such that the population of the ground state remains approximately 1 and that the population of the excited state is proportional to $\mathcal{E}^2$, the change in absorption is given by
\begin{align}
    \Delta \sigma (\omega, t) &= \sigma_\text{gs}(\omega) - \sigma_\text{exc}(\omega, t) \nonumber \\
    &\propto -\omega \mathcal{E}^2\sum_{n} \vert \bm{\mu}_n \vert^2  p_n(t)\delta(\omega-E_n)\,,
\end{align}
where $\bm{\mu}_n$ is the transition dipole moment from the ground state to excitonic state $n$ and $p_n(t)$ is the population of excitonic state $n$ at time $t$. 

\begin{figure}[ht]
\centering
\includegraphics[width=0.8\columnwidth]{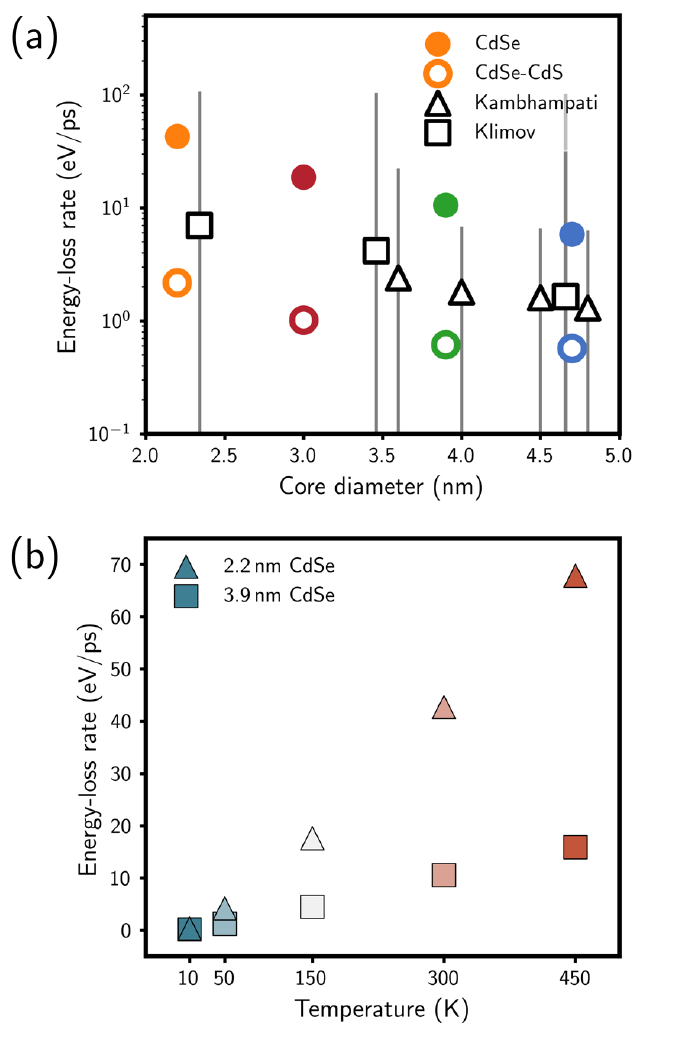}
\caption{(a) The energy loss rates calculated for CdSe and CdSe-CdS core-shell NCs. Core-shell structures all have 3 monolayers of shell. The black symbols correspond to measurements performed on CdSe NCs using transient absorption by Klimov \textit{et al.}\cite{Klimov1999} and state-resolved pump-probe spectroscopy by Cooney \textit{et al.}\cite{KambhampatiPRB2007} Vertical grey lines correspond to experimental error bars. (b) The energy loss rates calculated for 2.2\,nm CdSe and 3.9\,nm CdSe NCs show a linear dependence on temperature with a stronger temperature-dependence for the 2.2\,nm CdSe NC.}
\label{Fig5_lossRates}
\end{figure}

The change in absorption, $-\Delta \sigma (\omega, t)$, shows a fast decay of the $1P$ excitonic peak and a slower rise of the $1S$ ground excitonic peak (Fig.~S1). The dynamics of the rise of the $1S$ peak reflect those of hot exciton cooling. For each system, the rise dynamics were fit to an exponential function, and the extracted timescale was divided by the energy difference between the $1P$ and $1S$ excitonic peaks to yield an energy loss rate. The calculated energy loss rates are illustrated in Fig.~\ref{Fig5_lossRates}a along with those measured experimentally using transient absorption spectroscopy\cite{Klimov1999} and state-resolved pump-probe spectroscopy\cite{KambhampatiPRB2007} on wurtzite CdSe NCs. In agreement with experiment, our simulations show faster energy loss rates for smaller CdSe NCs. Smaller NCs have larger excitonic gaps due to quantum confinement and a smaller number of phonon modes, but they have stronger EXPC than larger NCs. Similarly, core-shell NCs, which have significantly weaker EXPC to lower-frequency acoustic modes,\cite{Jasrasaria2021} show energy loss rates that are an order of magnitude slower than those of bare cores.

\begin{figure*}[h!]
\centering
\includegraphics{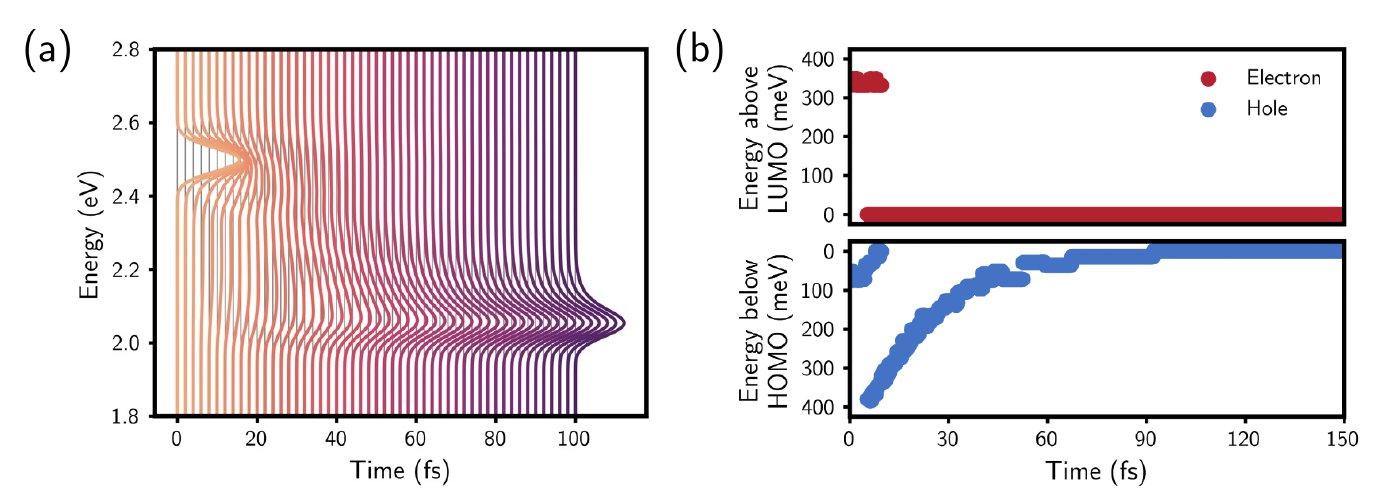}
\caption{(a) The calculated density of excitonic states of a 3.9\,nm CdSe NC scaled by the time-dependent population shows that hot exciton cooling occurs \textit{via} a cascade of relaxation events. (b) Projection of exciton cooling dynamics for the same CdSe NC onto a single-particle, electron/hole picture shows consistency with the Auger cooling mechanism. Hole relaxation is followed by an Auger-like process that leads to electron relaxation to the band edge and hole re-excitation. The hole then relaxes again.}
\label{Fig6_Mechanism}
\end{figure*}

While Fig.~\ref{Fig5_lossRates}a initially suggests that the calculated energy loss rates for CdSe NCs are larger than the measured values (but within the experimental error bars), those experiments use $\sim$100\,fs pulses that obscure the observation of dynamics between states with spectral overlap,\cite{Engel2013} like those measured here, and they measure NCs with very low photoluminescence quantum yields of around 1\%, where carrier trapping may lead to dynamics that complicate the hot exciton cooling process. Two-dimensional electronic spectroscopy (ES) measurements, which are able to clearly resolve the features corresponding to excitonic relaxation, on 3.5\,nm CdSe NCs show that hot exciton cooling from the higher-energy $1S$ excitonic peak to the ground excitonic peak occurs within $\sim$30\,fs,\cite{Engel2013} which is consistent with our findings. Furthermore, more recent two-dimensional ES experiments observe that cooling slows by an order of magnitude with the addition of a shell to a CdSe core,\cite{Kambhampati2023} in agreement with our calculated results.

While changing the size and composition of NCs is one avenue for tuning the cooling timescale, changing the temperature is another. The energy loss rates for 2.2\,nm CdSe and 3.9\,nm CdSe NCs simulated at different temperatures are illustrated in Fig.~\ref{Fig5_lossRates}b. For both NCs, the energy loss rates are $\sim 0.1$\,eV/ps at 10\,K, and they monotonically increase with temperature, as expected for phonon-mediated processes. While both systems show a linear relationship between energy loss rate and temperature, the rate for 2.2\,nm NC shows a stronger dependence on temperature than that of the 3.9\,nm NC. This steeper scaling with increasing temperature may be a result of the stronger quantum confinement in the 2.2\,nm NC, which leads to larger energy gaps between excitonic states. Thus, multiphonon processes at larger transition energies are more important. As those transition rates are very sensitive to temperature (Fig.~S2), the overall cooling process in smaller NCs has a stronger temperature dependence. Interestingly, as shown in Fig.~S3, the single-phonon-mediated cooling process also depends on temperature, but the dynamics converge above a threshold temperature. The threshold temperature is $\sim$300\,K for the 2.2\,nm CdSe NC while it is $\sim$50\,K for the 3.9\,nm CdSe. Again, this result may be due to the larger excitonic gaps in the smaller NC system. Note that our model given by Eq.~(\ref{eq:og_Hamiltonian}) includes EXPC to lowest order in the phonon mode coordinates and ignores higher-order terms (Duschinsky rotations), which may influence the hot exciton cooling process at higher temperatures.\cite{TommyLin2022}

\subsection{Mechanistic insight}

Finally, we investigate the mechanism underlying this ultrafast hot exciton cooling process. We calculated the density of excitonic states for a 3.9\,nm CdSe NC and scaled it by the time-dependent population, as illustrated in Fig.~\ref{Fig6_Mechanism}a. We see that cooling occurs \textit{via} a cascade of relaxation events through the manifold of excitonic states, as opposed to being dominated by a single or a few, higher-energy non-radiative transitions, as expected previously.\cite{Inoshita1992,Nozik2001,Schaller2005}

We wanted to understand the relationship between this multiphonon-mediated, hot exciton cooling mechanism (Fig.~\ref{Fig1_Mechanism}c) and the Auger-assisted cooling mechanism (Fig.~\ref{Fig1_Mechanism}b), which was first proposed to explain the breaking of the phonon bottleneck.\cite{Efros1995} To this end, we projected our simulated exciton cooling dynamics for a 3.9\,nm CdSe NC onto a single-particle picture of non-interacting electron-hole pair states, shown in Fig.~\ref{Fig6_Mechanism}b, and find that the Auger cooling mechanism emerges naturally from our excitonic dynamics. The hole quickly relaxes to the band edge \textit{via} multiphonon emission followed by electron relaxation by $\sim$400\,meV that results in hole re-excitation, and then the hole once again relaxes to the band edge by multiphonon emission. This result indicates that \textit{both} Coulomb-mediated electron-hole correlations, which are inherent in our formalism, \textit{and} multiphonon-mediated excitonic transitions are required to circumvent the phonon bottleneck and lead to ultrafast timescales of hot exciton cooling. These mechanistic insights are consistent for core-shell NCs, as illustrated in Fig.~S4.

\section{Conclusions}
Hot exciton cooling in confined semiconductor NCs involves rich physics, including electron-hole correlations, EXPC, and multiphonon-mediated nonradiative transitions---all of which are required to break the phonon bottleneck and enable fast relaxation of hot excitons to the band edge. We have developed the first atomistic theory that describes multiphonon-mediated exciton dynamics in NCs of experimentally relevant sizes. Our approach yields cooling timescales of tens of fs, which are consistent with measurements of similar systems. These ultrafast timescales are enabled by a cascade of multiphonon-mediated transitions between excitonic states that are relatively close in energy. These nonradiative transitions are made efficient by the large number of phonon modes in NCs that span a wide frequency range. The timescale of cooling is governed largely by the overall magnitude of EXPC, so that larger cores show slower relaxation, and core-shell NCs show relaxation that is slower by an order of magnitude.

Our approach provides fundamental insights to phonon-mediated exciton dynamics at the nanoscale, which differ significantly from those in molecular and bulk semiconductor systems. These simulations provide the first unified, microscopic theory for hot exciton cooling in nanoscale systems that addresses longstanding questions regarding the timescales and mechanisms of this process and that provides design principles for NCs with tuned EXPC and cooling timescales. The framework presented here is sufficiently general that it can be used to study timescales and mechanisms of exciton dephasing and carrier trapping. Furthermore, it can be used to investigate dynamics in NCs of different dimensionalities, such as in nanorods and nanoplatelets, and materials, including III-V, as long as EXPC is weak. Further elucidating the principles of phonon-mediated dynamics at the nanoscale is key to ultimately tuning these processes to realize novel phenomena in NC systems and NC-based applications with higher device efficiencies.

\section{Methods}

As described in previous work,\cite{Jasrasaria2021} CdSe and CdSe-CdS core-shell structures were optimized \textit{via} the LAMMPS molecular dynamics code\cite{Plimpton1995} using Stillinger-Weber interatomic potentials.\cite{Zhou2013} The outermost atomic monolayer was then removed and the subsequent monolayer was replaced by potentials representing the passivation layer.

Electronic structure calculations were performed using the semiempirical pseudopotential method.\cite{Wang1994, Wang1996, Rabani1999b} We used the filter-diagonalization technique to solve for single-particle electron and hole states near the band edges and then used these as input to the Bethe-Salpeter equation, which was solved to obtain correlated electron-hole pair states.\cite{Philbin2018} The exciton-phonon couplings were calculated within this framework,\cite{Jasrasaria2021} and phonon modes and frequencies were obtained by diagonalizing the dynamical matrix computed using the same Stillinger-Weber interatomic potentials.\cite{Zhou2013}

The correlation functions $\langle g_{n,m}(t)g_{m,n}(0)\rangle_\text{eq}$ in Eq.~(\ref{eq:rates_multiPhonon}) were evaluated within a harmonic approximation by sampling from a thermal distribution of bath coordinates and propagating classical trajectories. To facilitate convergence, we propagate and average trajectories to short times and then approximate the correlation functions as Gaussian functions. However, the timescales on which the correlation functions decay are not necessarily faster than those of the system dynamics, so we calculated non-Markovian relaxation rates. We also applied the standard quantum correction scheme to impose detailed balance.\cite{egorov1998semiclassical} Thus, the final transition rates are given by
\begin{align}
    \Gamma_{n\rightarrow m}(t) = &\frac{2}{\hbar^2 \big[1+e^{-\beta \hbar (\varepsilon_n -\varepsilon_m)}\big]}\times \nonumber \\
    \int_0^t d\tau \cos&\big((\varepsilon_n - \varepsilon_m)\tau/\hbar\big) \langle g_{n,m}(\tau)g_{m,n}(0)\rangle_\text{eq}\,.
\end{align}

We computed rates for all transitions and used them to build a kinetic master equation and propagate phonon-mediated exciton dynamics. Further details for all methods are provided in the Supporting Information.

\begin{suppinfo}
Procedure used to construct NC configurations; additional discussion regarding the implementation of the semi-empirical pseudopotential method, filter-diagonalization technique, Bethe-Salpeter equation, and EXPC matrix elements; detailed derivation and analysis of polaron transform and resulting Hamiltonian; illustration of calculated absorption spectra, temperature-dependent dynamics, and cooling mechanisms for core-shell NCs.
\end{suppinfo}

\begin{acknowledgement}
E.R. acknowledges support from the U.S. Department of Energy, Office of Science, Office of Basic Energy Sciences, Materials Sciences and Engineering Division, under Contract No. DE-AC02-05CH11231 within the Fundamentals of Semiconductor Nanowire Program (KCPY23). Methods used in this work were provided by the Center for Computational Study of Excited State Phenomena in Energy Materials (C2SEPEM), which is funded by the U.S. Department of Energy, Office of Science, Basic Energy Sciences, Materials Sciences and Engineering Division, \textit{via} Contract No. DE-AC02-05CH11231, as part of the Computational Materials Sciences Program. D.J. acknowledges the support of the Computational Science Graduate Fellowship from the U.S. Department of Energy under Grant No. DE-SC0019323.
\end{acknowledgement}

\begin{tocentry}
\begin{center}
\includegraphics{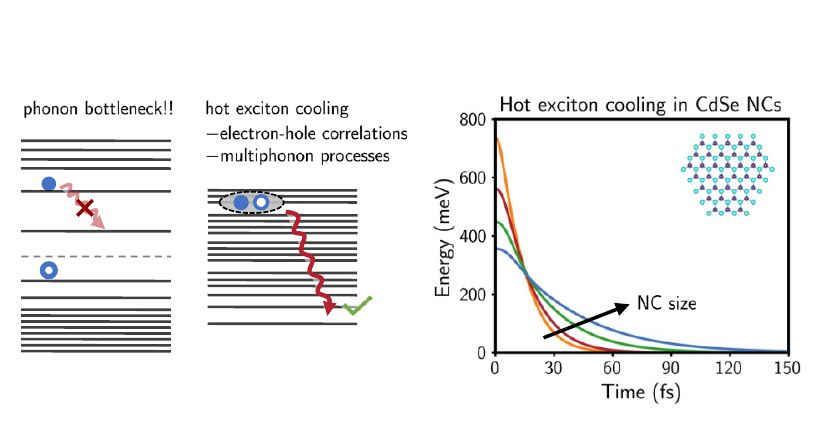}
\end{center}
\end{tocentry}


\begin{mcitethebibliography}{61}
\providecommand*\natexlab[1]{#1}
\providecommand*\mciteSetBstSublistMode[1]{}
\providecommand*\mciteSetBstMaxWidthForm[2]{}
\providecommand*\mciteBstWouldAddEndPuncttrue
  {\def\EndOfBibitem{\unskip.}}
\providecommand*\mciteBstWouldAddEndPunctfalse
  {\let\EndOfBibitem\relax}
\providecommand*\mciteSetBstMidEndSepPunct[3]{}
\providecommand*\mciteSetBstSublistLabelBeginEnd[3]{}
\providecommand*\EndOfBibitem{}
\mciteSetBstSublistMode{f}
\mciteSetBstMaxWidthForm{subitem}{(\alph{mcitesubitemcount})}
\mciteSetBstSublistLabelBeginEnd
  {\mcitemaxwidthsubitemform\space}
  {\relax}
  {\relax}

\bibitem[Lannoo \latin{et~al.}(1996)Lannoo, Delerue, and Allan]{Lannoo1996}
Lannoo,~M.; Delerue,~C.; Allan,~G. Theory of radiative and nonradiative
  transitions for semiconductor nanocrystals. \emph{J. Lumin.} \textbf{1996},
  \emph{70}, 170--184\relax
\mciteBstWouldAddEndPuncttrue
\mciteSetBstMidEndSepPunct{\mcitedefaultmidpunct}
{\mcitedefaultendpunct}{\mcitedefaultseppunct}\relax
\EndOfBibitem
\bibitem[Klimov(2006)]{klimov2006mechanisms}
Klimov,~V.~I. Mechanisms for photogeneration and recombination of multiexcitons
  in semiconductor nanocrystals: Implications for lasing and solar energy
  conversion. \emph{J. Phys. Chem. B} \textbf{2006}, \emph{110},
  16827--16845\relax
\mciteBstWouldAddEndPuncttrue
\mciteSetBstMidEndSepPunct{\mcitedefaultmidpunct}
{\mcitedefaultendpunct}{\mcitedefaultseppunct}\relax
\EndOfBibitem
\bibitem[Sheik-Bahae and Epstein(2007)Sheik-Bahae, and
  Epstein]{sheik2007optical}
Sheik-Bahae,~M.; Epstein,~R.~I. Optical refrigeration. \emph{Nat. Photon.}
  \textbf{2007}, \emph{1}, 693--699\relax
\mciteBstWouldAddEndPuncttrue
\mciteSetBstMidEndSepPunct{\mcitedefaultmidpunct}
{\mcitedefaultendpunct}{\mcitedefaultseppunct}\relax
\EndOfBibitem
\bibitem[Fomenko and Nesbitt(2008)Fomenko, and Nesbitt]{fomenko2008solution}
Fomenko,~V.; Nesbitt,~D.~J. Solution control of radiative and nonradiative
  lifetimes: A novel contribution to quantum dot blinking suppression.
  \emph{Nano Lett.} \textbf{2008}, \emph{8}, 287--293\relax
\mciteBstWouldAddEndPuncttrue
\mciteSetBstMidEndSepPunct{\mcitedefaultmidpunct}
{\mcitedefaultendpunct}{\mcitedefaultseppunct}\relax
\EndOfBibitem
\bibitem[Talapin \latin{et~al.}(2010)Talapin, Lee, Kovalenko, and
  Shevchenko]{Talapin2010}
Talapin,~D.~V.; Lee,~J.~S.; Kovalenko,~M.~V.; Shevchenko,~E.~V. {Prospects of
  colloidal nanocrystals for electronic and optoelectronic applications}.
  \emph{Chem. Rev.} \textbf{2010}, \emph{110}, 389--458\relax
\mciteBstWouldAddEndPuncttrue
\mciteSetBstMidEndSepPunct{\mcitedefaultmidpunct}
{\mcitedefaultendpunct}{\mcitedefaultseppunct}\relax
\EndOfBibitem
\bibitem[Hanifi \latin{et~al.}(2019)Hanifi, Bronstein, Koscher, Nett, Swabeck,
  Takano, Schwartzberg, Maserati, Vandewal, van~de Burgt, Salleo, and
  Alivisatos]{Hanifi2019}
Hanifi,~D.~A.; Bronstein,~N.~D.; Koscher,~B.~A.; Nett,~Z.; Swabeck,~J.~K.;
  Takano,~K.; Schwartzberg,~A.~M.; Maserati,~L.; Vandewal,~K.; van~de
  Burgt,~Y.; Salleo,~A.; Alivisatos,~A.~P. {Redefining near-unity luminescence
  in quantum dots with photothermal threshold quantum yield}. \emph{Science}
  \textbf{2019}, \emph{363}, 1199--1202\relax
\mciteBstWouldAddEndPuncttrue
\mciteSetBstMidEndSepPunct{\mcitedefaultmidpunct}
{\mcitedefaultendpunct}{\mcitedefaultseppunct}\relax
\EndOfBibitem
\bibitem[Peterson \latin{et~al.}(2014)Peterson, Cass, Harris, Edme, Sung, and
  Weiss]{Weiss2014}
Peterson,~M.~D.; Cass,~L.~C.; Harris,~R.~D.; Edme,~K.; Sung,~K.; Weiss,~E.~A.
  The role of ligands in determining the exciton relaxation dynamics in
  semiconductor quantum dots. \emph{Annu. Rev. Phys. Chem.} \textbf{2014},
  \emph{65}, 317--339\relax
\mciteBstWouldAddEndPuncttrue
\mciteSetBstMidEndSepPunct{\mcitedefaultmidpunct}
{\mcitedefaultendpunct}{\mcitedefaultseppunct}\relax
\EndOfBibitem
\bibitem[Melnychuk and Guyot-Sionnest(2021)Melnychuk, and
  Guyot-Sionnest]{PGS2021}
Melnychuk,~C.; Guyot-Sionnest,~P. Multicarrier dynamics in quantum dots.
  \emph{Chem. Rev.} \textbf{2021}, \emph{121}, 2325--2372\relax
\mciteBstWouldAddEndPuncttrue
\mciteSetBstMidEndSepPunct{\mcitedefaultmidpunct}
{\mcitedefaultendpunct}{\mcitedefaultseppunct}\relax
\EndOfBibitem
\bibitem[von~der Linde and Lambrich(1979)von~der Linde, and
  Lambrich]{Lambrich1979}
von~der Linde,~D.; Lambrich,~R. Direct measurement of hot-electron relaxation
  by picosecond spectroscopy. \emph{Phys. Rev. Lett.} \textbf{1979}, \emph{42},
  1090--1093\relax
\mciteBstWouldAddEndPuncttrue
\mciteSetBstMidEndSepPunct{\mcitedefaultmidpunct}
{\mcitedefaultendpunct}{\mcitedefaultseppunct}\relax
\EndOfBibitem
\bibitem[Pugnet \latin{et~al.}(1981)Pugnet, Collet, and Cornet]{Pugnet1981}
Pugnet,~M.; Collet,~J.; Cornet,~A. Cooling of hot electron-hole plasmas in the
  presence of screened electron-phonon interactions. \emph{Solid State Commun.}
  \textbf{1981}, \emph{38}, 531--536\relax
\mciteBstWouldAddEndPuncttrue
\mciteSetBstMidEndSepPunct{\mcitedefaultmidpunct}
{\mcitedefaultendpunct}{\mcitedefaultseppunct}\relax
\EndOfBibitem
\bibitem[Prabhu \latin{et~al.}(1995)Prabhu, Vengurlekar, Roy, and
  Shah]{Prabhu1995}
Prabhu,~S.~S.; Vengurlekar,~A.~S.; Roy,~S.~K.; Shah,~J. Nonequilibrium dynamics
  of hot carriers and hot phonons in CdSe and GaAs. \emph{Phys. Rev. B}
  \textbf{1995}, \emph{51}, 14233--14246\relax
\mciteBstWouldAddEndPuncttrue
\mciteSetBstMidEndSepPunct{\mcitedefaultmidpunct}
{\mcitedefaultendpunct}{\mcitedefaultseppunct}\relax
\EndOfBibitem
\bibitem[Kambhampati(2011)]{Kambhampati2011}
Kambhampati,~P. Hot exciton relaxation dynamics in semiconductor quantum dots:
  radiationless transitions on the nanoscale. \emph{J. Phys. Chem. C}
  \textbf{2011}, \emph{115}, 22089--22109\relax
\mciteBstWouldAddEndPuncttrue
\mciteSetBstMidEndSepPunct{\mcitedefaultmidpunct}
{\mcitedefaultendpunct}{\mcitedefaultseppunct}\relax
\EndOfBibitem
\bibitem[Knowles \latin{et~al.}(2011)Knowles, McArthur, and Weiss]{Knowles2011}
Knowles,~K.~E.; McArthur,~E.~A.; Weiss,~E.~A. A multi-timescale map of
  radiative and nonradiative decay pathways for excitons in CdSe quantum dots.
  \emph{ACS Nano} \textbf{2011}, \emph{5}, 2026--2035\relax
\mciteBstWouldAddEndPuncttrue
\mciteSetBstMidEndSepPunct{\mcitedefaultmidpunct}
{\mcitedefaultendpunct}{\mcitedefaultseppunct}\relax
\EndOfBibitem
\bibitem[Jasrasaria \latin{et~al.}(2022)Jasrasaria, Weinberg, Philbin, and
  Rabani]{Jasrasaria2022_JCP}
Jasrasaria,~D.; Weinberg,~D.; Philbin,~J.~P.; Rabani,~E. Simulations of
  nonradiative processes in semiconductor nanocrystals. \emph{J. Chem. Phys.}
  \textbf{2022}, \emph{157}, 020901\relax
\mciteBstWouldAddEndPuncttrue
\mciteSetBstMidEndSepPunct{\mcitedefaultmidpunct}
{\mcitedefaultendpunct}{\mcitedefaultseppunct}\relax
\EndOfBibitem
\bibitem[Nozik(2001)]{Nozik2001}
Nozik,~A.~J. {Spectroscopy and hot electron relaxation dynamics in
  semiconductor quantum wells and quantum dots}. \emph{Annu. Rev. Phys. Chem.}
  \textbf{2001}, 193--231\relax
\mciteBstWouldAddEndPuncttrue
\mciteSetBstMidEndSepPunct{\mcitedefaultmidpunct}
{\mcitedefaultendpunct}{\mcitedefaultseppunct}\relax
\EndOfBibitem
\bibitem[Inoshita and Sakaki(1992)Inoshita, and Sakaki]{Inoshita1992}
Inoshita,~T.; Sakaki,~H. Electron relaxation in a quantum dot: Significance of
  multiphonon processes. \emph{Phys. Rev. B} \textbf{1992}, \emph{46},
  7260--7263\relax
\mciteBstWouldAddEndPuncttrue
\mciteSetBstMidEndSepPunct{\mcitedefaultmidpunct}
{\mcitedefaultendpunct}{\mcitedefaultseppunct}\relax
\EndOfBibitem
\bibitem[Gfroerer \latin{et~al.}(1996)Gfroerer, Sturge, Kash, Yater, Plaut,
  Lin, Florez, Harbison, Das, and Lebrun]{Gfroerer1996}
Gfroerer,~T.~H.; Sturge,~M.~D.; Kash,~K.; Yater,~J.~A.; Plaut,~A.~S.; Lin,~P.
  S.~D.; Florez,~L.~T.; Harbison,~J.~P.; Das,~S.~R.; Lebrun,~L. Slow relaxation
  of excited states in strain-induced quantum dots. \emph{Phys. Rev. B}
  \textbf{1996}, \emph{53}, 16474--16480\relax
\mciteBstWouldAddEndPuncttrue
\mciteSetBstMidEndSepPunct{\mcitedefaultmidpunct}
{\mcitedefaultendpunct}{\mcitedefaultseppunct}\relax
\EndOfBibitem
\bibitem[Yu \latin{et~al.}(1996)Yu, Lycett, Roberts, and Murray]{Haiping1996}
Yu,~H.; Lycett,~S.; Roberts,~C.; Murray,~R. Time resolved study of
  self-assembled InAs quantum dots. \emph{Appl. Phys. Lett.} \textbf{1996},
  \emph{69}, 4087--4089\relax
\mciteBstWouldAddEndPuncttrue
\mciteSetBstMidEndSepPunct{\mcitedefaultmidpunct}
{\mcitedefaultendpunct}{\mcitedefaultseppunct}\relax
\EndOfBibitem
\bibitem[Heitz \latin{et~al.}(1997)Heitz, Veit, Ledentsov, Hoffmann, Bimberg,
  Ustinov, Kop'ev, and Alferov]{Heitz1997}
Heitz,~R.; Veit,~M.; Ledentsov,~N.~N.; Hoffmann,~A.; Bimberg,~D.;
  Ustinov,~V.~M.; Kop'ev,~P.~S.; Alferov,~Z.~I. Energy relaxation by
  multiphonon processes in InAs/GaAs quantum dots. \emph{Phys. Rev. B}
  \textbf{1997}, \emph{56}, 10435--10445\relax
\mciteBstWouldAddEndPuncttrue
\mciteSetBstMidEndSepPunct{\mcitedefaultmidpunct}
{\mcitedefaultendpunct}{\mcitedefaultseppunct}\relax
\EndOfBibitem
\bibitem[Guyot-Sionnest \latin{et~al.}(1999)Guyot-Sionnest, Shim, Matranga, and
  Hines]{GS1999}
Guyot-Sionnest,~P.; Shim,~M.; Matranga,~C.; Hines,~M. Intraband relaxation in
  CdSe quantum dots. \emph{Phys. Rev. B} \textbf{1999}, \emph{60},
  R2181--R2184\relax
\mciteBstWouldAddEndPuncttrue
\mciteSetBstMidEndSepPunct{\mcitedefaultmidpunct}
{\mcitedefaultendpunct}{\mcitedefaultseppunct}\relax
\EndOfBibitem
\bibitem[Sosnowski \latin{et~al.}(1998)Sosnowski, Norris, Jiang, Singh, Kamath,
  and Bhattacharya]{Sosnowski1998}
Sosnowski,~T.~S.; Norris,~T.~B.; Jiang,~H.; Singh,~J.; Kamath,~K.;
  Bhattacharya,~P. Rapid carrier relaxation in
  ${\mathrm{In}}_{0.4}{\mathrm{Ga}}_{0.6}\mathrm{A}\mathrm{s}/\mathrm{G}\mathrm{a}\mathrm{A}\mathrm{s}$
  quantum dots characterized by differential transmission spectroscopy.
  \emph{Phys. Rev. B} \textbf{1998}, \emph{57}, R9423--R9426\relax
\mciteBstWouldAddEndPuncttrue
\mciteSetBstMidEndSepPunct{\mcitedefaultmidpunct}
{\mcitedefaultendpunct}{\mcitedefaultseppunct}\relax
\EndOfBibitem
\bibitem[Mukai and Sugawara(1998)Mukai, and Sugawara]{Mukai1998}
Mukai,~K.; Sugawara,~M. Slow carrier relaxation among sublevels in annealed
  self-formed {InGaAs}/{GaAs} quantum dots. \emph{Jpn. J. Appl. Phys.}
  \textbf{1998}, \emph{37}, 5451--5456\relax
\mciteBstWouldAddEndPuncttrue
\mciteSetBstMidEndSepPunct{\mcitedefaultmidpunct}
{\mcitedefaultendpunct}{\mcitedefaultseppunct}\relax
\EndOfBibitem
\bibitem[Pandey and Guyot-Sionnest(2008)Pandey, and Guyot-Sionnest]{Pandey2008}
Pandey,~A.; Guyot-Sionnest,~P. Slow electron cooling in colloidal quantum dots.
  \emph{Science} \textbf{2008}, \emph{322}, 929--932\relax
\mciteBstWouldAddEndPuncttrue
\mciteSetBstMidEndSepPunct{\mcitedefaultmidpunct}
{\mcitedefaultendpunct}{\mcitedefaultseppunct}\relax
\EndOfBibitem
\bibitem[Asahi(1997)]{asahi1997self}
Asahi,~H. Self-organized quantum wires and dots in III--V semiconductors.
  \emph{Adv. Mater.} \textbf{1997}, \emph{9}, 1019--1026\relax
\mciteBstWouldAddEndPuncttrue
\mciteSetBstMidEndSepPunct{\mcitedefaultmidpunct}
{\mcitedefaultendpunct}{\mcitedefaultseppunct}\relax
\EndOfBibitem
\bibitem[Klimov and McBranch(1998)Klimov, and McBranch]{klimov1998femtosecond}
Klimov,~V.~I.; McBranch,~D.~W. Femtosecond 1P-to-1S electron relaxation in
  strongly confined semiconductor nanocrystals. \emph{Phys. Rev. Lett.}
  \textbf{1998}, \emph{80}, 4028--4031\relax
\mciteBstWouldAddEndPuncttrue
\mciteSetBstMidEndSepPunct{\mcitedefaultmidpunct}
{\mcitedefaultendpunct}{\mcitedefaultseppunct}\relax
\EndOfBibitem
\bibitem[Klimov \latin{et~al.}(1999)Klimov, McBranch, Leatherdale, and
  Bawendi]{Klimov1999}
Klimov,~V.~I.; McBranch,~D.~W.; Leatherdale,~C.~A.; Bawendi,~M.~G. Electron and
  hole relaxation pathways in semiconductor quantum dots. \emph{Phys. Rev. B}
  \textbf{1999}, \emph{60}, 13740--13749\relax
\mciteBstWouldAddEndPuncttrue
\mciteSetBstMidEndSepPunct{\mcitedefaultmidpunct}
{\mcitedefaultendpunct}{\mcitedefaultseppunct}\relax
\EndOfBibitem
\bibitem[Klimov \latin{et~al.}(2000)Klimov, Mikhailovsky, McBranch,
  Leatherdale, and Bawendi]{Klimov2000a}
Klimov,~V.~I.; Mikhailovsky,~A.~A.; McBranch,~D.~W.; Leatherdale,~C.~A.;
  Bawendi,~M.~G. Mechanisms for intraband energy relaxation in semiconductor
  quantum dots: The role of electron-hole interactions. \emph{Phys. Rev. B}
  \textbf{2000}, \emph{61}, R13349--R13352\relax
\mciteBstWouldAddEndPuncttrue
\mciteSetBstMidEndSepPunct{\mcitedefaultmidpunct}
{\mcitedefaultendpunct}{\mcitedefaultseppunct}\relax
\EndOfBibitem
\bibitem[Schaller \latin{et~al.}(2005)Schaller, Pietryga, Goupalov, Petruska,
  Ivanov, and Klimov]{Schaller2005}
Schaller,~R.~D.; Pietryga,~J.~M.; Goupalov,~S.~V.; Petruska,~M.~A.;
  Ivanov,~S.~A.; Klimov,~V.~I. Breaking the phonon bottleneck in semiconductor
  nanocrystals via multiphonon emission induced by intrinsic nonadiabatic
  interactions. \emph{Phys. Rev. Lett.} \textbf{2005}, \emph{95}, 196401\relax
\mciteBstWouldAddEndPuncttrue
\mciteSetBstMidEndSepPunct{\mcitedefaultmidpunct}
{\mcitedefaultendpunct}{\mcitedefaultseppunct}\relax
\EndOfBibitem
\bibitem[Harbold \latin{et~al.}(2005)Harbold, Du, Krauss, Cho, Murray, and
  Wise]{Harbold2005}
Harbold,~J.~M.; Du,~H.; Krauss,~T.~D.; Cho,~K.-S.; Murray,~C.~B.; Wise,~F.~W.
  Time-resolved intraband relaxation of strongly confined electrons and holes
  in colloidal PbSe nanocrystals. \emph{Phys. Rev. B} \textbf{2005}, \emph{72},
  195312\relax
\mciteBstWouldAddEndPuncttrue
\mciteSetBstMidEndSepPunct{\mcitedefaultmidpunct}
{\mcitedefaultendpunct}{\mcitedefaultseppunct}\relax
\EndOfBibitem
\bibitem[Cooney \latin{et~al.}(2007)Cooney, Sewall, Dias, Sagar, Anderson, and
  Kambhampati]{KambhampatiPRB2007}
Cooney,~R.~R.; Sewall,~S.~L.; Dias,~E.~A.; Sagar,~D.~M.; Anderson,~K. E.~H.;
  Kambhampati,~P. Unified picture of electron and hole relaxation pathways in
  semiconductor quantum dots. \emph{Phys. Rev. B} \textbf{2007}, \emph{75},
  245311\relax
\mciteBstWouldAddEndPuncttrue
\mciteSetBstMidEndSepPunct{\mcitedefaultmidpunct}
{\mcitedefaultendpunct}{\mcitedefaultseppunct}\relax
\EndOfBibitem
\bibitem[Cooney \latin{et~al.}(2007)Cooney, Sewall, Anderson, Dias, and
  Kambhampati]{KambhampatiPRL2007}
Cooney,~R.~R.; Sewall,~S.~L.; Anderson,~K. E.~H.; Dias,~E.~A.; Kambhampati,~P.
  Breaking the phonon bottleneck for holes in semiconductor quantum dots.
  \emph{Phys. Rev. Lett.} \textbf{2007}, \emph{98}, 177403\relax
\mciteBstWouldAddEndPuncttrue
\mciteSetBstMidEndSepPunct{\mcitedefaultmidpunct}
{\mcitedefaultendpunct}{\mcitedefaultseppunct}\relax
\EndOfBibitem
\bibitem[Kharchenko and Rosen(1996)Kharchenko, and Rosen]{Kharchenko1996}
Kharchenko,~V.; Rosen,~M. Auger relaxation processes in semiconductor
  nanocrystals and quantum wells. \emph{J. Lumin.} \textbf{1996}, \emph{70},
  158--169\relax
\mciteBstWouldAddEndPuncttrue
\mciteSetBstMidEndSepPunct{\mcitedefaultmidpunct}
{\mcitedefaultendpunct}{\mcitedefaultseppunct}\relax
\EndOfBibitem
\bibitem[Efros \latin{et~al.}(1995)Efros, Kharchenko, and Rosen]{Efros1995}
Efros,~A.~L.; Kharchenko,~V.; Rosen,~M. Breaking the phonon bottleneck in
  nanometer quantum dots: Role of Auger-like processes. \emph{Solid State
  Commun.} \textbf{1995}, \emph{93}, 281--284\relax
\mciteBstWouldAddEndPuncttrue
\mciteSetBstMidEndSepPunct{\mcitedefaultmidpunct}
{\mcitedefaultendpunct}{\mcitedefaultseppunct}\relax
\EndOfBibitem
\bibitem[Hendry \latin{et~al.}(2006)Hendry, Koeberg, Wang, Zhang,
  de~Mello~Doneg\'a, Vanmaekelbergh, and Bonn]{Hendry2006}
Hendry,~E.; Koeberg,~M.; Wang,~F.; Zhang,~H.; de~Mello~Doneg\'a,~C.;
  Vanmaekelbergh,~D.; Bonn,~M. Direct observation of electron-to-hole energy
  transfer in CdSe quantum dots. \emph{Phys. Rev. Lett.} \textbf{2006},
  \emph{96}, 057408\relax
\mciteBstWouldAddEndPuncttrue
\mciteSetBstMidEndSepPunct{\mcitedefaultmidpunct}
{\mcitedefaultendpunct}{\mcitedefaultseppunct}\relax
\EndOfBibitem
\bibitem[Philbin and Rabani(2018)Philbin, and Rabani]{Philbin2018}
Philbin,~J.~P.; Rabani,~E. {Electron-hole correlations govern Auger
  recombination in nanostructures}. \emph{Nano Lett.} \textbf{2018}, \emph{18},
  7889--7895\relax
\mciteBstWouldAddEndPuncttrue
\mciteSetBstMidEndSepPunct{\mcitedefaultmidpunct}
{\mcitedefaultendpunct}{\mcitedefaultseppunct}\relax
\EndOfBibitem
\bibitem[Guyot-Sionnest \latin{et~al.}(2005)Guyot-Sionnest, Wehrenberg, and
  Yu]{GS2005}
Guyot-Sionnest,~P.; Wehrenberg,~B.; Yu,~D. Intraband relaxation in CdSe
  nanocrystals and the strong influence of the surface ligands. \emph{J. Chem.
  Phys.} \textbf{2005}, \emph{123}, 074709\relax
\mciteBstWouldAddEndPuncttrue
\mciteSetBstMidEndSepPunct{\mcitedefaultmidpunct}
{\mcitedefaultendpunct}{\mcitedefaultseppunct}\relax
\EndOfBibitem
\bibitem[Xu \latin{et~al.}(2002)Xu, Mikhailovsky, Hollingsworth, and
  Klimov]{Klimov2002}
Xu,~S.; Mikhailovsky,~A.~A.; Hollingsworth,~J.~A.; Klimov,~V.~I. Hole intraband
  relaxation in strongly confined quantum dots: Revisiting the ``phonon
  bottleneck'' problem. \emph{Phys. Rev. B} \textbf{2002}, \emph{65},
  045319\relax
\mciteBstWouldAddEndPuncttrue
\mciteSetBstMidEndSepPunct{\mcitedefaultmidpunct}
{\mcitedefaultendpunct}{\mcitedefaultseppunct}\relax
\EndOfBibitem
\bibitem[Wang \latin{et~al.}(2003)Wang, Califano, Zunger, and
  Franceschetti]{Wang2003}
Wang,~L.-W.; Califano,~M.; Zunger,~A.; Franceschetti,~A. {Pseudopotential
  theory of Auger processes in CdSe quantum dots}. \emph{Phys. Rev. Lett.}
  \textbf{2003}, \emph{91}, 056404\relax
\mciteBstWouldAddEndPuncttrue
\mciteSetBstMidEndSepPunct{\mcitedefaultmidpunct}
{\mcitedefaultendpunct}{\mcitedefaultseppunct}\relax
\EndOfBibitem
\bibitem[Kilina \latin{et~al.}(2009)Kilina, Kilin, and Prezhdo]{Kilina2009}
Kilina,~S.~V.; Kilin,~D.~S.; Prezhdo,~O.~V. {Breaking the phonon bottleneck in
  PbSe and CdSe quantum dots: Time-domain density functional theory of charge
  carrier relaxation}. \emph{ACS Nano} \textbf{2009}, \emph{3}, 93--99\relax
\mciteBstWouldAddEndPuncttrue
\mciteSetBstMidEndSepPunct{\mcitedefaultmidpunct}
{\mcitedefaultendpunct}{\mcitedefaultseppunct}\relax
\EndOfBibitem
\bibitem[Prezhdo(2009)]{Oleg2009}
Prezhdo,~O.~V. Photoinduced dynamics in semiconductor quantum dots: Insights
  from time-domain ab initio studies. \emph{Acc. Chem. Res.} \textbf{2009},
  \emph{42}, 2005--2016\relax
\mciteBstWouldAddEndPuncttrue
\mciteSetBstMidEndSepPunct{\mcitedefaultmidpunct}
{\mcitedefaultendpunct}{\mcitedefaultseppunct}\relax
\EndOfBibitem
\bibitem[Zeng and He(2021)Zeng, and He]{Zeng2021}
Zeng,~T.; He,~Y. {Ab initio modeling of phonon-assisted relaxation of electrons
  and excitons in semiconductor nanocrystals for multiexciton generation}.
  \emph{Phys. Rev. B} \textbf{2021}, \emph{103}, 1--15\relax
\mciteBstWouldAddEndPuncttrue
\mciteSetBstMidEndSepPunct{\mcitedefaultmidpunct}
{\mcitedefaultendpunct}{\mcitedefaultseppunct}\relax
\EndOfBibitem
\bibitem[Pandey and Guyot-Sionnest(2010)Pandey, and
  Guyot-Sionnest]{pandey2010hot}
Pandey,~A.; Guyot-Sionnest,~P. Hot electron extraction from colloidal quantum
  dots. \emph{J. Phys. Chem. Lett.} \textbf{2010}, \emph{1}, 45--47\relax
\mciteBstWouldAddEndPuncttrue
\mciteSetBstMidEndSepPunct{\mcitedefaultmidpunct}
{\mcitedefaultendpunct}{\mcitedefaultseppunct}\relax
\EndOfBibitem
\bibitem[Jasrasaria and Rabani(2021)Jasrasaria, and Rabani]{Jasrasaria2021}
Jasrasaria,~D.; Rabani,~E. Interplay of surface and interior modes in
  exciton-phonon coupling at the nanoscale. \emph{Nano Lett.} \textbf{2021},
  \emph{21}, 8741--8748\relax
\mciteBstWouldAddEndPuncttrue
\mciteSetBstMidEndSepPunct{\mcitedefaultmidpunct}
{\mcitedefaultendpunct}{\mcitedefaultseppunct}\relax
\EndOfBibitem
\bibitem[Jasrasaria and Rabani(2022)Jasrasaria, and
  Rabani]{Jasrasaria2022erratum}
Jasrasaria,~D.; Rabani,~E. Correction to interplay of surface and interior
  modes in exciton--phonon coupling at the nanoscale. \emph{Nano Lett.}
  \textbf{2022}, \emph{22}, 8033--8034\relax
\mciteBstWouldAddEndPuncttrue
\mciteSetBstMidEndSepPunct{\mcitedefaultmidpunct}
{\mcitedefaultendpunct}{\mcitedefaultseppunct}\relax
\EndOfBibitem
\bibitem[Zhou \latin{et~al.}(2013)Zhou, Ward, Martin, {Van Swol}, Cruz-Campa,
  and Zubia]{Zhou2013}
Zhou,~X.~W.; Ward,~D.~K.; Martin,~J.~E.; {Van Swol},~F.~B.; Cruz-Campa,~J.~L.;
  Zubia,~D. {Stillinger-Weber potential for the II-VI elements
  Zn-Cd-Hg-S-Se-Te}. \emph{Phys. Rev. B} \textbf{2013}, \emph{88}, 085309\relax
\mciteBstWouldAddEndPuncttrue
\mciteSetBstMidEndSepPunct{\mcitedefaultmidpunct}
{\mcitedefaultendpunct}{\mcitedefaultseppunct}\relax
\EndOfBibitem
\bibitem[Sewall \latin{et~al.}(2006)Sewall, Cooney, Anderson, Dias, and
  Kambhampati]{Sewall2006}
Sewall,~S.~L.; Cooney,~R.~R.; Anderson,~K.~E.; Dias,~E.~A.; Kambhampati,~P.
  State-to-state exciton dynamics in semiconductor quantum dots. \emph{Phys.
  Rev. B} \textbf{2006}, \emph{74}, 235328\relax
\mciteBstWouldAddEndPuncttrue
\mciteSetBstMidEndSepPunct{\mcitedefaultmidpunct}
{\mcitedefaultendpunct}{\mcitedefaultseppunct}\relax
\EndOfBibitem
\bibitem[Nitzan and Press(2006)Nitzan, and Press]{nitzan2006chemical}
Nitzan,~A.; Press,~O.~U. \emph{Chemical Dynamics in Condensed Phases:
  Relaxation, Transfer and Reactions in Condensed Molecular Systems}; Oxford
  Graduate Texts; OUP Oxford, 2006\relax
\mciteBstWouldAddEndPuncttrue
\mciteSetBstMidEndSepPunct{\mcitedefaultmidpunct}
{\mcitedefaultendpunct}{\mcitedefaultseppunct}\relax
\EndOfBibitem
\bibitem[Zimanyi and Silbey(2012)Zimanyi, and
  Silbey]{SilbeyPolaronTransform2012}
Zimanyi,~E.~N.; Silbey,~R.~J. Theoretical description of quantum effects in
  multi-chromophoric aggregates. \emph{Philos. Trans. R. Soc. A} \textbf{2012},
  \emph{370}, 3620--3637\relax
\mciteBstWouldAddEndPuncttrue
\mciteSetBstMidEndSepPunct{\mcitedefaultmidpunct}
{\mcitedefaultendpunct}{\mcitedefaultseppunct}\relax
\EndOfBibitem
\bibitem[Xu and Cao(2016)Xu, and Cao]{Xu2016}
Xu,~D.; Cao,~J. Non-canonical distribution and non-equilibrium transport beyond
  weak system-bath coupling regime: A polaron transformation approach.
  \emph{Front. Phys.} \textbf{2016}, \emph{11}, 110308\relax
\mciteBstWouldAddEndPuncttrue
\mciteSetBstMidEndSepPunct{\mcitedefaultmidpunct}
{\mcitedefaultendpunct}{\mcitedefaultseppunct}\relax
\EndOfBibitem
\bibitem[Franchini \latin{et~al.}(2021)Franchini, Reticcioli, Setvin, and
  Diebold]{Franchini2021}
Franchini,~C.; Reticcioli,~M.; Setvin,~M.; Diebold,~U. Polarons in materials.
  \emph{Nat. Rev. Mater.} \textbf{2021}, \emph{6}, 560--586\relax
\mciteBstWouldAddEndPuncttrue
\mciteSetBstMidEndSepPunct{\mcitedefaultmidpunct}
{\mcitedefaultendpunct}{\mcitedefaultseppunct}\relax
\EndOfBibitem
\bibitem[Englman and Jortner(1970)Englman, and Jortner]{EnglmanJortner1970}
Englman,~R.; Jortner,~J. The energy gap law for radiationless transitions in
  large molecules. \emph{Mol. Phys.} \textbf{1970}, \emph{18}, 145--164\relax
\mciteBstWouldAddEndPuncttrue
\mciteSetBstMidEndSepPunct{\mcitedefaultmidpunct}
{\mcitedefaultendpunct}{\mcitedefaultseppunct}\relax
\EndOfBibitem
\bibitem[Eshet \latin{et~al.}(2013)Eshet, Gr{\"{u}}nwald, and
  Rabani]{Eshet2013}
Eshet,~H.; Gr{\"{u}}nwald,~M.; Rabani,~E. {The electronic structure of CdSe/CdS
  core/shell seeded nanorods: Type-I or quasi-type-II?} \emph{Nano Lett.}
  \textbf{2013}, \emph{13}, 5880--5885\relax
\mciteBstWouldAddEndPuncttrue
\mciteSetBstMidEndSepPunct{\mcitedefaultmidpunct}
{\mcitedefaultendpunct}{\mcitedefaultseppunct}\relax
\EndOfBibitem
\bibitem[Griffin \latin{et~al.}(2013)Griffin, Ithurria, Dolzhnikov, Linkin,
  Talapin, and Engel]{Engel2013}
Griffin,~G.~B.; Ithurria,~S.; Dolzhnikov,~D.~S.; Linkin,~A.; Talapin,~D.~V.;
  Engel,~G.~S. Two-dimensional electronic spectroscopy of CdSe nanoparticles at
  very low pulse power. \emph{J. Chem. Phys.} \textbf{2013}, \emph{138},
  014705\relax
\mciteBstWouldAddEndPuncttrue
\mciteSetBstMidEndSepPunct{\mcitedefaultmidpunct}
{\mcitedefaultendpunct}{\mcitedefaultseppunct}\relax
\EndOfBibitem
\bibitem[Brosseau \latin{et~al.}(2022)Brosseau, Geuchies, Jasrasaria, Houtepen,
  Rabani, and Kambhampati]{Kambhampati2023}
Brosseau,~P.~J.; Geuchies,~J.~J.; Jasrasaria,~D.; Houtepen,~A.~J.; Rabani,~E.;
  Kambhampati,~P. New ultrafast hole relaxation channels in quantum dots
  revealed by two-dimensional electronic spectroscopy. \emph{In review.}
  \textbf{2022}, \relax
\mciteBstWouldAddEndPunctfalse
\mciteSetBstMidEndSepPunct{\mcitedefaultmidpunct}
{}{\mcitedefaultseppunct}\relax
\EndOfBibitem
\bibitem[Lin \latin{et~al.}(2022)Lin, Jasrasaria, Yoo, Bawendi, Utzat, and
  Rabani]{TommyLin2022}
Lin,~K.; Jasrasaria,~D.; Yoo,~J.~J.; Bawendi,~M.; Utzat,~H.; Rabani,~E. Theory
  of photoluminescence spectral line shapes of semiconductor nanocrystals.
  \emph{arXiv:2212.06323} \textbf{2022}, \relax
\mciteBstWouldAddEndPunctfalse
\mciteSetBstMidEndSepPunct{\mcitedefaultmidpunct}
{}{\mcitedefaultseppunct}\relax
\EndOfBibitem
\bibitem[Plimpton(1995)]{Plimpton1995}
Plimpton,~S. {Fast parallel algorithms for short-range molecular dynamics}.
  \emph{J. Comput. Phys.} \textbf{1995}, \emph{117}, 1--19\relax
\mciteBstWouldAddEndPuncttrue
\mciteSetBstMidEndSepPunct{\mcitedefaultmidpunct}
{\mcitedefaultendpunct}{\mcitedefaultseppunct}\relax
\EndOfBibitem
\bibitem[Wang and Zunger(1994)Wang, and Zunger]{Wang1994}
Wang,~L.~W.; Zunger,~A. {Electronic structure pseudopotential calculations of
  large ($\sim$1000 atoms) Si quantum dots}. \emph{J. Phys. Chem.}
  \textbf{1994}, \emph{98}, 2158--2165\relax
\mciteBstWouldAddEndPuncttrue
\mciteSetBstMidEndSepPunct{\mcitedefaultmidpunct}
{\mcitedefaultendpunct}{\mcitedefaultseppunct}\relax
\EndOfBibitem
\bibitem[Wang and Zunger(1996)Wang, and Zunger]{Wang1996}
Wang,~L.-W.; Zunger,~A. {Pseudopotential calculations of nanoscale CdSe quantum
  dots}. \emph{Phys. Rev. B} \textbf{1996}, \emph{53}, 9579--9582\relax
\mciteBstWouldAddEndPuncttrue
\mciteSetBstMidEndSepPunct{\mcitedefaultmidpunct}
{\mcitedefaultendpunct}{\mcitedefaultseppunct}\relax
\EndOfBibitem
\bibitem[Rabani \latin{et~al.}(1999)Rabani, Hetenyi, Berne, and
  Brus]{Rabani1999b}
Rabani,~E.; Hetenyi,~B.; Berne,~B.~J.; Brus,~L.~E. {Electronic properties of
  CdSe nanocrystals in the absence and presence of a dielectric medium}.
  \emph{J. Chem. Phys.} \textbf{1999}, \emph{110}, 5355--5369\relax
\mciteBstWouldAddEndPuncttrue
\mciteSetBstMidEndSepPunct{\mcitedefaultmidpunct}
{\mcitedefaultendpunct}{\mcitedefaultseppunct}\relax
\EndOfBibitem
\bibitem[Egorov and Skinner(1998)Egorov, and Skinner]{egorov1998semiclassical}
Egorov,~S.; Skinner,~J. Semiclassical approximations to quantum time
  correlation functions. \emph{Chem. Phys. Lett.} \textbf{1998}, \emph{293},
  469--476\relax
\mciteBstWouldAddEndPuncttrue
\mciteSetBstMidEndSepPunct{\mcitedefaultmidpunct}
{\mcitedefaultendpunct}{\mcitedefaultseppunct}\relax
\EndOfBibitem
\end{mcitethebibliography}

\providecommand{\latin}[1]{#1}
\makeatletter
\providecommand{\doi}
  {\begingroup\let\do\@makeother\dospecials
  \catcode`\{=1 \catcode`\}=2 \doi@aux}
\providecommand{\doi@aux}[1]{\endgroup\texttt{#1}}
\makeatother
\providecommand*\mcitethebibliography{\thebibliography}
\csname @ifundefined\endcsname{endmcitethebibliography}
  {\let\endmcitethebibliography\endthebibliography}{}

\end{document}


\newpage
\section{Nanostructure configurations}

Nanostructure configurations for CdSe cores were obtained by cleaving a wurtzite crystal with a lattice constant of bulk wurtzite CdSe ($a=4.30\,$\AA, $c=a\sqrt{\frac{8}{3}}$) such that all atoms are bonded to at least two other atoms. The structure was then minimized with respect to a Stillinger-Weber interatomic potential, which was previously parameterized for II-VI structures,\cite{Zhou2013} using the conjugate gradient descent minimization implemented in LAMMPS.\cite{Plimpton1995} The outermost monolayer of atoms was removed and each semiconductor atom remaining on the surface was then replaced with a ligand pseudopotential, which represents the passivation layer.\cite{Rabani1999b} The ligand pseudopotentials push up the energies of localized states to prevent the creation of mid-gap trap states. For CdSe-CdS core-shell structures, four monolayers of CdS were added to the CdSe core structure before geometry minimization, monolayer removal, and the addition of the passivation layer such that each final core-shell structure has three monolayers of CdS shell. The configurations and diameters for each CdSe and CdSe-CdS core-shell configuration studied here are collected in Tables~\ref{tab:CdSe_NCs} and \ref{tab:CdSe-CdS_NCs}, respectively.

\begin{center}
\begin{table}
\caption{CdSe nanocrystal configuration details.\label{tab:CdSe_NCs}}
\begin{centering}
\begin{tabular}{cc}
\hline 
Configuration & Diameter (nm)\tabularnewline
\hline 
$\text{Cd}_{93}\text{Se}_{93}$ & 2.2\tabularnewline
$\text{Cd}_{222}\text{Se}_{222}$ & 3.0\tabularnewline
$\text{Cd}_{435}\text{Se}_{435}$ & 3.9\tabularnewline
$\text{Cd}_{753}\text{Se}_{753}$ & 4.7\tabularnewline
\hline 
\end{tabular}
\par\end{centering}
\end{table}
\par\end{center}

\begin{center}
\begin{table}
\caption{CdSe-CdS core-shell nanocrystal configuration details. All structures
have 3 monolayers of CdS shell.\label{tab:CdSe-CdS_NCs}}
\centering{}%
\begin{tabular}{ccc}
\hline 
Configuration & Core diameter (nm) & Total diameter (nm)\tabularnewline
\hline 
$\text{Cd}_{102}\text{Se}_{102}$--$\text{Cd}_{651}\text{Se}_{651}$ & 2.2 & 4.6\tabularnewline
$\text{Cd}_{243}\text{Se}_{243}$--$\text{Cd}_{954}\text{Se}_{954}$ & 3.0 & 5.4\tabularnewline
$\text{Cd}_{462}\text{Se}_{462}$--$\text{Cd}_{1326}\text{Se}_{1326}$ & 3.9 & 6.2\tabularnewline
$\text{Cd}_{798}\text{Se}_{798}$--$\text{Cd}_{1749}\text{Se}_{1749}$ & 4.7 & 7.1\tabularnewline
\hline 
\end{tabular}
\end{table}
\par\end{center}

\section{Semi-empirical pseudopotential method and filter diagonalization technique}

All calculations were performed within the semi-empirical pseudopotential method for CdSe and CdSe-CdS structures.\cite{Rabani1999b} The local, screened pseudopotentials were fit to reproduce bulk band structures, effective masses, and absolute deformation potentials of wurtzite and zincblende CdSe and CdS. The functional form for the local pseudopotential for atom type $\mu$ in momentum-space is given by
\begin{equation}
    \hat{\tilde{v}}_\mu (q) = a_0^\mu \big[ 1 + a_4^\mu (\text{Tr}\epsilon_\mu) + a_5^\mu (\text{Tr}\epsilon_\mu)^3 \big] \frac{q^2 - a_1^\mu}{a_2^\mu \exp(a_3^\mu q^2) - 1}\,,
\label{eq:pp}
\end{equation}
where $\text{Tr}\epsilon_\mu$ is the local strain tensor of atom $\mu$, which is calculated as the volume of the tetrahedron formed by the nearest neighbors of atom $\mu$. The parameters $\{a^\mu\}$ for Cd, Se, and S are collected in Table~\ref{tab:parameters}. Additional details about pseudopotential fitting are given in Ref.~\citen{Jasrasaria2022_JCP}.

\begin{center}
\begin{table}[ht]
\caption{Pseudopotential parameters for Cd, Se, and S. All parameters are given
in atomic units.}
\begin{centering}
\begin{tabular}{ccccccc}
\hline 
 & $a_{0}$ & $a_{1}$ & $a_{2}$ & $a_{3}$ & $a_{4}$ & $a_{5}$\tabularnewline
\hline 
Cd & $-$31.4518 & 1.3890 & $-$0.0502 & 1.6603 & 0.0586 & 0\tabularnewline
Se & 8.4921 & 4.3513 & 1.3600 & 0.3227 & 0.1746 & $-$33\tabularnewline
S & 7.6697 & 4.5192 & 1.3456 & 0.3035 & 0.2087 & 0\tabularnewline
\end{tabular}
\par\end{centering}
\label{tab:parameters}
\end{table}
\par\end{center}

Calculations were implemented on real-space grids with spacings less than 0.8\,a.u., which is sufficient to converge the results. The filter-diagonalization technique, which filters electron and hole states near target energies set as the LUMO and HOMO energies, was used to obtain quasiparticle states $\phi(\bm{r}_e)$ and $\phi(\bm{r}_h)$ at the conduction and valence band edges, respectively. These states are eigenstates of the single-particle Hamiltonian.

\section{Bethe-Salpeter equation}

Excitonic (\textit{i.e.}, correlated electron-hole pair) states were represented as a linear combination of non-interacting electron-hole pair states:
\begin{equation}
    \psi_n (\bm{r}_e, \bm{r}_h) = \sum_{ai} c_{a,i}^n \phi_a(\bm{r}_e) \phi_i(\bm{r}_h)\,,
\label{eqn:BSE_psi}
\end{equation}
where $a$ refers to single-particle electron (unoccupied) states, and $i$ refers to single-particle hole (occupied) states. The coefficients $\{ c_{a,i}^n \}$ were obtained by solving the Bethe-Salpeter equation (BSE) within the static screening approximation. The number of single-particle states used in the BSE was $\sim$100-200, with larger numbers of states needed for larger systems. All BSE calculations converged the energies for all of the excitonic states involved in the cooling process studied here. Additional details regarding the BSE are given in Ref.~\citen{Philbin2018}.

\section{Exciton-phonon coupling}

The exciton-phonon coupling (EXPC) was expanded to first order in the phonon mode coordinates, as seen in Eq.~(1) of the main text. The exciton-nuclear matrix elements are given by
\begin{equation}
    V_{n,m}^{\mu k} \equiv \bigg\langle \psi_n \bigg\vert \bigg( \frac{\partial \hat{v}(\bm{r})}{\partial R_{\mu k}} \bigg)_{\boldsymbol{R}_0} \bigg\vert \psi_m \bigg\rangle\,,
\end{equation}
where $\vert \psi_{n} \rangle$ is the state of exciton $n$, $\hat{v}(\bm{r}) = \sum_\mu \hat{v}_\mu(\boldsymbol{r})$ is the sum over atomic pseudopotentials given in Eq.~(\ref{eq:pp}), $R_{\mu k}$ is the position of atom $\mu$ in the $k\in \{x,y,z\}$ direction, and $\boldsymbol{R}_0$ is the equilibrium configuration of the NC. Given the form of the excitonic wave function in Eq.~(\ref{eqn:BSE_psi}), we can reduce the calculation of these matrix elements to a simpler form in terms of the BSE coefficients, $\{ c_{a,i}^n \}$, and real-space single-particle wavefunctions, $\phi(\bm{r})$:
\begin{equation}
    V_{n,m}^{\mu k} = \sum_{abi} c_{a,i}^n c_{b,i}^m v_{ab,\mu}^\prime (R_{\mu k}) - \sum_{aij} c_{a,i}^n c_{a,j}^m v_{ij,\mu}^\prime (R_{\mu k})\,,
\label{eqn:ExPh_ElHoleChannels}
\end{equation}
where
\begin{equation}
    v_{rs,\mu}^\prime = \int d\boldsymbol{r} \phi^{*}_r (\boldsymbol{r}) \frac{\partial \hat{v}(\bm{r})}{\partial R_{\mu k}} \phi_s (\boldsymbol{r})\,.
\label{eqn:single-ElHoleChannels}
\end{equation}

The exciton-nuclear matrix elements can be transformed to phonon mode coordinates using the eigenvectors of the dynamical matrix:
\begin{equation}
    V_{n,m}^\alpha = \sum_{\mu k} \frac{1}{\sqrt{m_\mu}} e_{\alpha,\mu k}^{-1} V_{n,m}^{\mu k}\,,    
\end{equation}
where $e_{\alpha, \mu k}$ is the $\mu k$ element of the $\alpha$ eigenvector of the dynamical matrix (\textit{i.e.}, mass-weighted Hessian with respect to the Stillinger-Weber potential\cite{Zhou2013}), and $m_{\mu}$ is the mass of atom $\mu$. The diagonal matrix elements, $V_{n,n}^\alpha$, describe the renormalization of the energy of exciton $n$ through its interaction with phonon mode $\alpha$, and the off-diagonal matrix elements, $V_{n,m}^\alpha$, describe the interaction of excitons $n$ and $m$ through the absorption or emission of a phonon of mode $\alpha$. Additional details regarding the calculation of EXPCs are given in Ref.~\citen{Jasrasaria2021}.

\section{Polaron-transformed Hamiltonian}

We use a model Hamiltonian to describes a manifold of excitonic states and phonons that are coupled to first order in the atomic displacements, which is given by\cite{Jasrasaria2021}
\begin{align}
\bm{H} & =\bm{H}_{S}+H_{B}+\bm{V}_{SB}\\
& =\sum_{n}E_{n}\left|\psi_{n}\right\rangle \left\langle \psi_{n}\right|+\sum_{\alpha}\hbar\omega_{\alpha}b_{\alpha}^{\dagger}b_{\alpha}+\sum_{\alpha nm}\sqrt{\frac{\hbar}{2\omega_{\alpha}}}V_{n,m}^{\alpha}\left|\psi_{n}\right\rangle \left\langle \psi_{m}\right|\left(b_{\alpha}^{\dagger}+b_{\alpha}\right)\\
& =\sum_{n}E_{n}\left|\psi_{n}\right\rangle \left\langle \psi_{n}\right|+\frac{1}{2}\sum_{\alpha}\left[p_{\alpha}^{2}+\omega_{\alpha}^{2}q_{\alpha}^{2}\right]+\sum_{\alpha nm}V_{n,m}^{\alpha}\left|\psi_{n}\right\rangle \left\langle \psi_{m}\right|q_{\alpha}\,. \label{eq:og_Hamiltonian2}
\end{align}
Note that operators with degrees of freedom in the excitonic subspace are bolded, and Eq.~(\ref{eq:og_Hamiltonian2}) makes use of the relationships between the Bosonic creation and annihilation operators and the position and momentum operators. We use mass-weighted coordinates.

We perform a unitary polaron transformation:\cite{nitzan2006chemical, Xu2016}
\begin{align}
\tilde{\bm{H}} & =e^{\bm{S}}\bm{H} e^{-\bm{S}}\,,\\
& =\exp\left(-\frac{i}{\hbar}\sum_{\alpha}\omega_{\alpha}^{-2}p_{\alpha}\bm{V}_{D}^{\alpha}\right)\bm{H}\exp\left(+\frac{i}{\hbar}\sum_{\alpha}\omega_{\alpha}^{-2}p_{\alpha}\bm{V}_{D}^{\alpha}\right)\,, \label{eqn:s11}
\end{align}
where we have defined the diagonal of a system-subspace operator,
\textbf{$\bm{A}$}, as \textbf{$\bm{A}_{D}$}, so
that
\begin{equation}
\bm{V}_{D}^{\alpha}\equiv\sum_{kl}V_{k,l}^{\alpha}\delta_{k,l}\left|\psi_{k}\right\rangle \left\langle \psi_{l}\right|\,.
\end{equation}

Because the matrices $\left[\bm{V}_{D}^{\alpha},\bm{V}_{D}^{\beta}\right]=0$ for all $\alpha$ and $\beta$, the exponential of a sum in Eq.~(\ref{eqn:s11}) can be written as a product of exponential functions:
\begin{equation}
e^{\bm{S}} =\prod_{\alpha}\exp\left(\sqrt{\frac{\omega_{\alpha}}{2\hbar}}\omega_{\alpha}^{-2}\left(b_{\alpha}^{\dagger}-b_{\alpha}\right)\bm{V}_{D}^{\alpha}\right)\,.
\end{equation}
This unitary operator is simply the set of position shift operators along modes $\alpha$.

For an operator in the system subspace, $\bm{A}$, we will define
\begin{equation}
\tilde{\bm{A}}\equiv e^{\bm{S}}\bm{A}e^{-\bm{S}}\,.
\end{equation}
Note that $\tilde{\bm{A}}$ is an operator in both the system and bath subspaces. Given these relationships, we can transform each term in the Hamiltonian in Eq.~(\ref{eq:og_Hamiltonian2}).

The polaron-transformed system Hamiltonian is given by
\begin{align}
\tilde{\bm{H}}_{S} & =e^{\bm{S}}\bm{H}_{S}e^{-\bm{S}}\\
& =e^{\bm{S}}\left(\sum_{n}E_{n}\left|\psi_{n}\right\rangle \left\langle \psi_{n}\right|\right)e^{-\bm{S}}\\
& =\bm{H}_{S}\,.
\end{align}
Because $\left[\bm{H}_{S},\bm{V}_{D}^{\alpha}\right]=0$,
the system Hamiltonian is unchanged.

The transform of the bath Hamiltonian is given by
\begin{align}
\tilde{\bm{H}}_{B} & =e^{\bm{S}}H_{B}e^{-\bm{S}}\\
& =\frac{1}{2}\sum_{\alpha}p_{\alpha}^{2}+\frac{1}{2}\sum_{\alpha}\omega_{\alpha}^{2}\left(q_{\alpha}-\frac{\bm{V}_{D}^{\alpha}}{\omega_{\alpha}^{2}}\right)^{2}\\
& =H_{B}-\sum_{\alpha}\bm{V}_{D}^{\alpha}q_{\alpha}+\frac{1}{2}\sum_{\alpha}\omega_{\alpha}^{-2}\left(\bm{V}_{D}^{\alpha}\right)^{2}\,.
\end{align}

Finally, the transformed EXPC is given by
\begin{align}
\tilde{\bm{V}}_{SB}= & e^{\bm{S}}\bm{V}_{SB}e^{-\bm{S}}\\
= & e^{\bm{S}}\left[\sum_{\alpha}\sqrt{\frac{\hbar}{2\omega_{\alpha}}}\bm{V}^{\alpha}\left(b_{\alpha}^{\dagger}+b_{\alpha}\right)\right]e^{-\bm{S}}\\
= & \frac{1}{2}\sum_{\alpha}\left[\tilde{\bm{V}}^{\alpha}\left(q_{\alpha}-\frac{i}{\omega_{\alpha}}p_{\alpha}\right)+\left(q_{\alpha}+\frac{i}{\omega_{\alpha}}p_{\alpha}\right)\tilde{\bm{V}}^{\alpha}-\omega_{\alpha}^{-2}\left(\tilde{\bm{V}}^{\alpha}\bm{V}_{D}^{\alpha}+\bm{V}_{D}^{\alpha}\tilde{\bm{V}}^{\alpha}\right)\right]\,.
\end{align}

Combining these results gives
\begin{align}
\tilde{\bm{H}}= & \bm{H}_{S}+\frac{1}{2}\sum_{\alpha}\omega_{\alpha}^{-2}\left(\bm{V}_{D}^{\alpha}\right)^{2}+H_{B}-\sum_{\alpha}\bm{V}_{D}^{\alpha}x_{\alpha}\nonumber \\
& +\frac{1}{2}\sum_{\alpha}\left[\tilde{\bm{V}}^{\alpha}\left(q_{\alpha}-\frac{i}{\omega_{\alpha}}p_{\alpha}\right)+\left(q_{\alpha}+\frac{i}{\omega_{\alpha}}p_{\alpha}\right)\tilde{\bm{V}}^{\alpha}-\omega_{\alpha}^{-2}\left(\tilde{\bm{V}}^{\alpha}\bm{V}_{D}^{\alpha}+\bm{V}_{D}^{\alpha}\tilde{\bm{V}}^{\alpha}\right)\right]\\
= & \sum_{n}\left(E_{n}-\lambda_{n}\right)\left|\psi_{n}\right\rangle \left\langle \psi_{n}\right|+H_{B}\nonumber \\
& +\frac{1}{2}\sum_{\alpha n\neq m}\left[\tilde{V}_{n,m}^{\alpha}\left(q_{\alpha}-\frac{i}{\omega_{\alpha}}p_{\alpha}\right)+\left(q_{\alpha}+\frac{i}{\omega_{\alpha}}p_{\alpha}\right)\tilde{V}_{n,m}^{\alpha}\right]\left|\psi_{n}\right\rangle \left\langle \psi_{m}\right| \nonumber \\
& -\sum_{n\neq m}\tilde{\lambda}_{nm}\left|\psi_{n}\right\rangle \left\langle \psi_{m}\right|\,,
\end{align}
where
\begin{align}
\lambda_{n} & \equiv\frac{1}{2}\sum_{\alpha}\omega_{\alpha}^{-2}\left(V_{n,n}^{\alpha}\right)^{2}\\
\tilde{\lambda}_{nm} & \equiv\frac{1}{2}\sum_{\alpha}\omega_{\alpha}^{-2}\tilde{V}_{n,m}^{\alpha}\left(V_{m,m}^{\alpha}+V_{n,n}^{\alpha}\right)\\
\tilde{V}_{n,m}^{\alpha} & \equiv\exp\left(-\frac{i}{\hbar}\sum_{\alpha}\omega_{\alpha}^{-2}p_{\alpha}V_{n,n}^{\alpha}\right)V_{n,m}^{\alpha}\exp\left(+\frac{i}{\hbar}\sum_{\alpha}\omega_{\alpha}^{-2}p_{\alpha}V_{m,m}^{\alpha}\right)\,.
\end{align}
Replacing the quantum position and momentum operators with their classical counterparts so that they commute with one another simplifies the above equation to
\begin{align}
\tilde{\bm{H}}= & \sum_{n}\left(E_{n}-\lambda_{n}\right)\left|\psi_{n}\right\rangle \left\langle \psi_{n}\right|+H_{B}\nonumber \\
& +\sum_{\alpha n\neq m}\tilde{V}_{n,m}^{\alpha}\left|\psi_{n}\right\rangle \left\langle \psi_{m}\right|q_{\alpha}-\sum_{n\neq m}\tilde{\lambda}_{nm}\left|\psi_{n}\right\rangle \left\langle \psi_{m}\right|\,.
\label{eq:polaron_Hamiltonian}
\end{align}

This Hamiltonian lends itself to a simple perturbation theory, where the first line on the right-hand side of Eq.~(\ref{eq:polaron_Hamiltonian}) describes the system, where the energies of excitonic states are renormalized by the reorganization energies, and the bath. The second lines describe the coupling between states through the bath. The first term of the system-bath coupling, $\sum_{\alpha}\tilde{V}_{n,m}^{\alpha}\left|\psi_{n}\right\rangle \left\langle \psi_{m}\right| q_\alpha$, mirrors the EXPC in Eq.~(\ref{eq:og_Hamiltonian2}) but the EXPC matrix elements are dressed by exponential functions, which can be interpreted as a rescaling of the EXPC by a set of mode shifts that are related to the Franck-Condon factors of the system.

These exponential functions give rise to multiphonon processes in the Fermi's golden rule rate, which can be seen by performing a Taylor expansion of the exponential functions in $\tilde{V}_{n,m}^{\alpha}q_{\alpha}$. The zeroth order expansion yields the form of the original EXPC to first order in the phonon mode coordinates, which gives rise to single-phonon processes. Higher order terms of the expansion describe the EXPC to higher orders of the phonon mode coordinates, accounting for multiphonon processes:
\begin{align}
\tilde{V}_{n,m}^{\alpha}q_{\alpha}= & V_{n,m}^{\alpha}q_{\alpha}\nonumber \\
 & -\frac{i}{\hbar}\sum_{\beta}\omega_{\beta}^{-2}p_{\beta}\left(V_{n,n}^{\beta}-V_{m,m}^{\beta}\right)V_{n,m}^{\alpha}q_{\alpha}\nonumber \\
 & +\frac{1}{\hbar^{2}}\sum_{\beta\gamma}\omega_{\beta}^{-2}\omega_{\gamma}^{-2}p_{\beta}p_{\gamma}V_{n,n}^{\beta}V_{m,m}^{\gamma}V_{n,m}^{\alpha}q_{\alpha}\nonumber \\
 & +\dots\,.
\end{align}

\section{Calculating changes in absorption spectra}

We simulate changes in the absorption spectrum of a system initially excited to the $1P$ excitonic state as it relaxes to the $1S$ ground excitonic state.
The ground state absorption is given by\cite{Hilborn1982}
\begin{equation}
    \sigma_\text{gs}(\omega) \propto \sum_{n} \vert \bm{\mu}_n \vert^2 \omega \delta(\omega-E_n)\,,
\end{equation}
where $\bm{\mu}_n$ is the transition dipole moment from the ground state to excitonic state $n$:
\begin{equation}
    \bm{\mu}_n = \sum_{ai} c_{a,i}^n \int d\bm{r} \phi_a (\bm{r}) \bm{r} \phi_i (\bm{r})\,.
\end{equation}
Here, $\{c_{a,i}^n\}$ are the Bethe-Salpeter coefficients introduced in Eq.~(\ref{eqn:BSE_psi}), and $\phi(\bm{r})$ are the real-space single-particle wavefunctions.
Assuming that the electric field, $\mathcal{E}$, is weak such that the population of the ground state remains approximately 1 and that population of the excited state is proportional to $\mathcal{E}^2$, the change in absorption is given by
\begin{align}
    \Delta \sigma (\omega, t) &= \sigma_\text{gs}(\omega) - \sigma_\text{exc}(\omega, t) \nonumber \\
    &\propto -\omega \mathcal{E}^2\sum_{n} \vert \bm{\mu}_n \vert^2  p_n(t)\delta(\omega-E_n)\,,
\end{align}
where $p_n(t)$ is the population of excitonic state $n$ at time $t$. 

\begin{figure}[h!]
\centering
\includegraphics[width=0.5\textwidth]{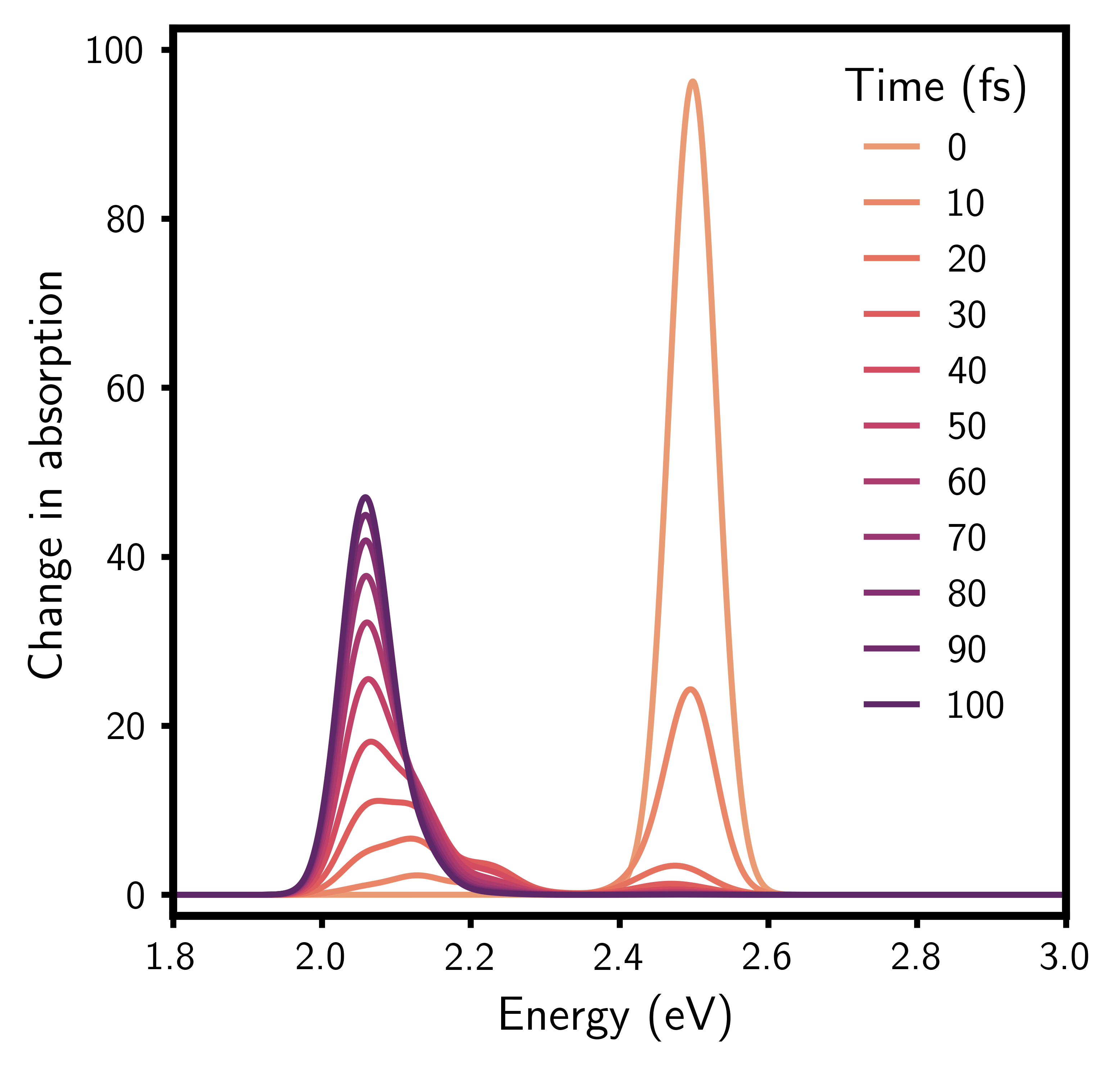}
\caption{The calculated change in absorption, $-\Delta \sigma(\omega, t)$, for a 3.9\,nm CdSe NC. The rise of the $1S$ ground excitonic peak reflects the process of hot exciton cooling.}
\label{FigS1_abs}
\end{figure}

The change in absorption, $-\Delta \sigma (\omega, t)$, of a 3.9\,nm CdSe NC is illustrated in Fig.~\ref{FigS1_abs}. The fast decay of the $1P$ excitonic peak and a slower rise of the $1S$ ground excitonic peak reflect the dynamics of hot exciton cooling.

\newpage
\section{Temperature-dependent hot exciton cooling}

\begin{figure}[h!]
\centering
\includegraphics[width=\textwidth]{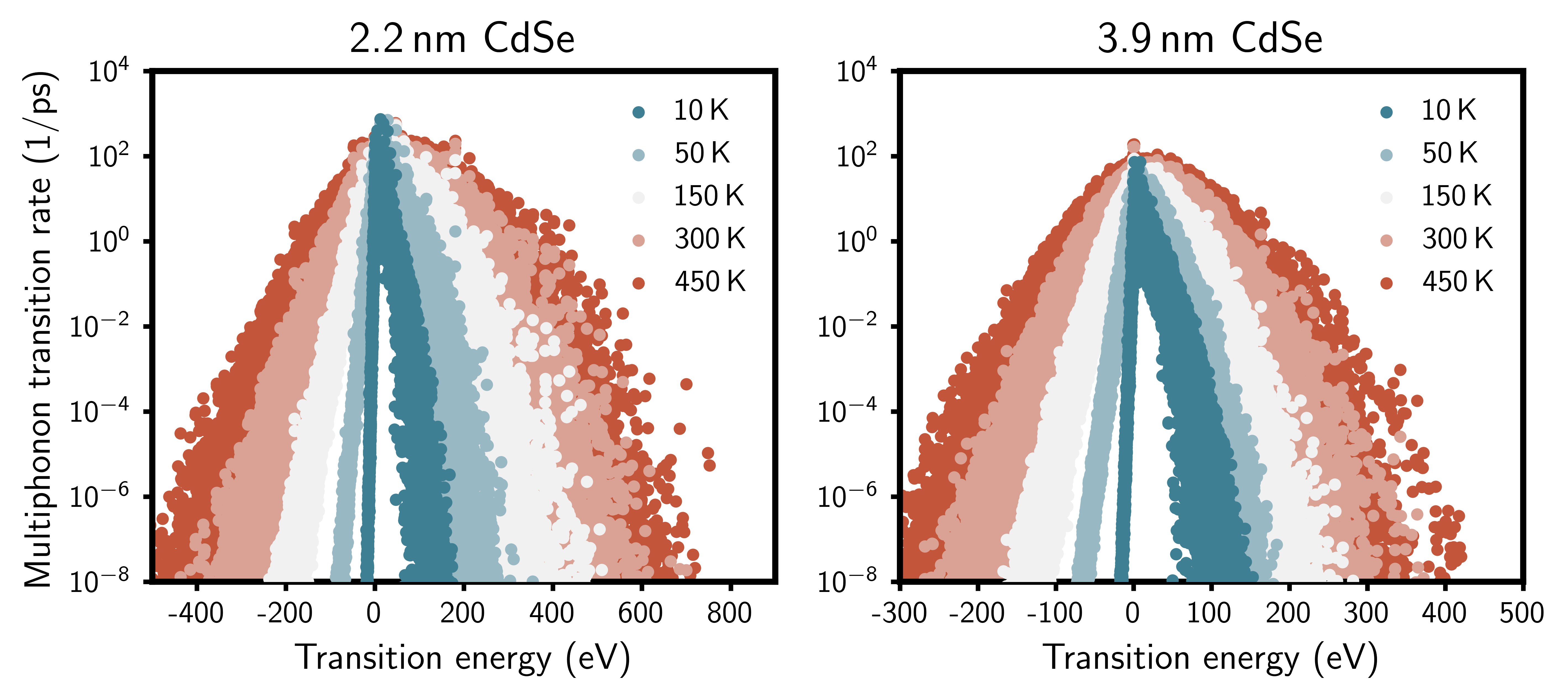}
\caption{Multiphonon exciton transition rates calculated for a 2.2\,nm CdSe NC (left) and 3.9\,nm CdSe NC (right) at different temperatures. The magnitude of the transition rates, especially for transition energies between $\sim30-100$\,meV, are very sensitive to temperature.}
\label{FigS2_tempRates}
\end{figure}

\begin{figure}[h!]
\centering
\includegraphics[width=\textwidth]{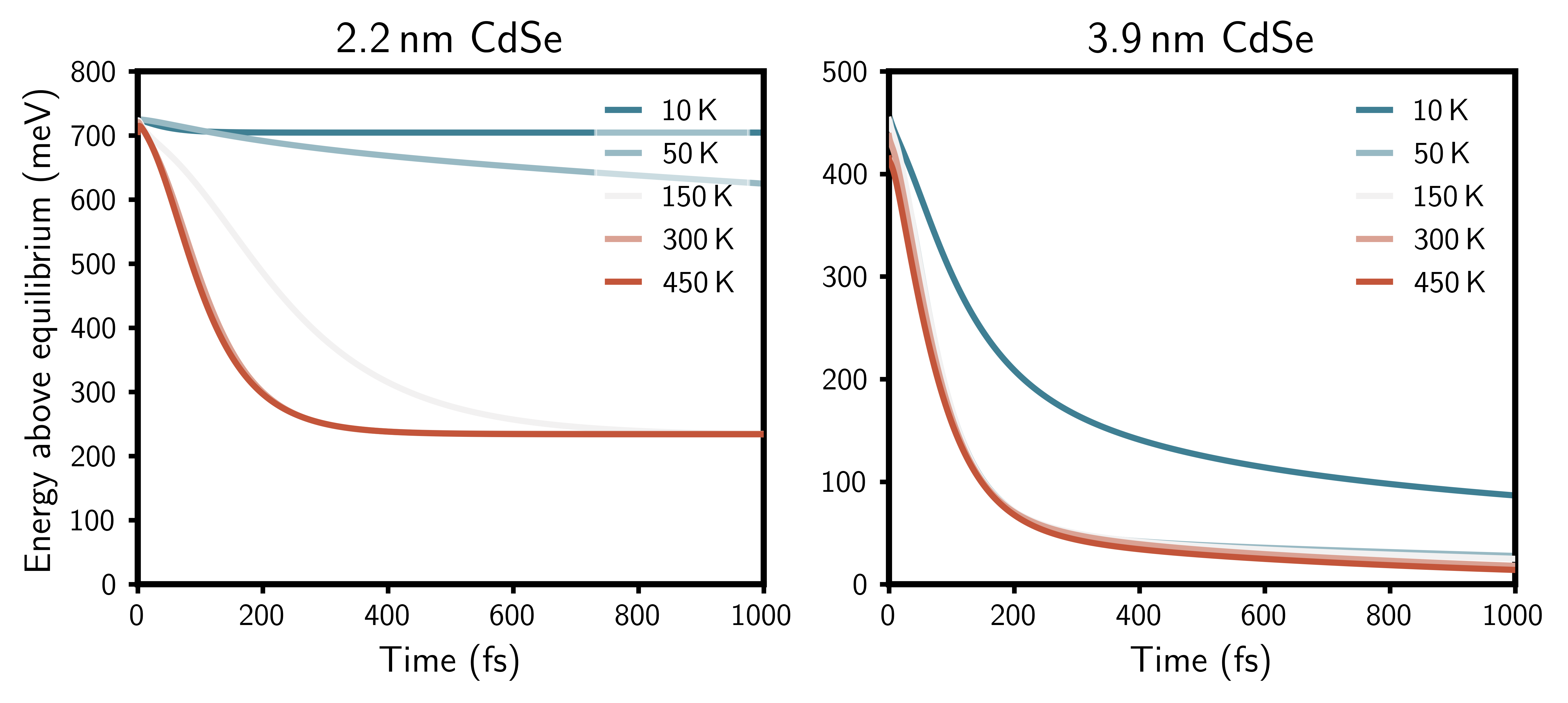}
\caption{Single-phonon-mediated hot exciton cooling simulated for a 2.2\,nm CdSe NC (left) and 3.9\,nm CdSe NC (right) at different temperatures. The temperature-dependent dynamics converge by 300\,K for the 2.2\,nm CdSe NC, while they converge by 50\,K for the 3.9\,nm CdSe NC. The phonon bottleneck persists at all temperatures within this single-phonon framework.}
\label{FigS3_tempDyn}
\end{figure}

\section{Hot exciton cooling mechanism of core-shell NCs}

\begin{figure}[h!]
\centering
\includegraphics[width=\textwidth]{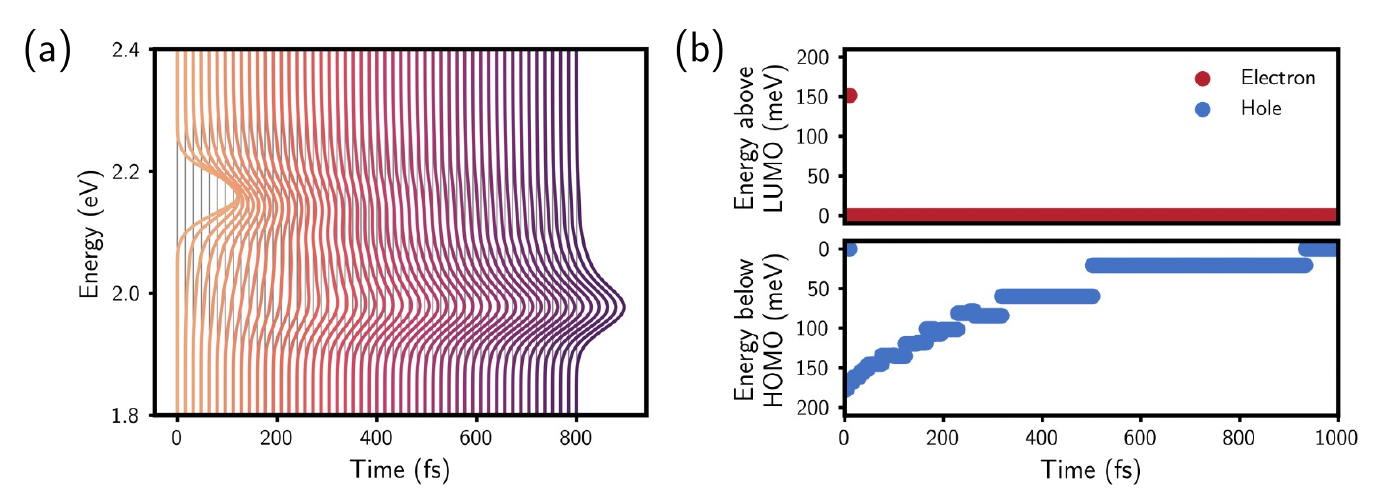}
\caption{(a) The calculated density of excitonic states of a 3.9\,nm CdSe -- 3\,ML CdS core-shell NC scaled by the time-dependent populations shows that hot exciton cooling occurs \textit{via} a cascade of relaxation events. (b) Projection of the exciton cooling dynamics for a 3.9\,nm CdSe -- 3\,ML CdS core-shell NC onto a single-particle, electron/hole picture shows consistency with the Auger cooling mechanism. These mechanistic insights are consistent with those of cooling in bare CdSe cores.}
\label{FigS4_mech}
\end{figure}


\providecommand{\latin}[1]{#1}
\makeatletter
\providecommand{\doi}
  {\begingroup\let\do\@makeother\dospecials
  \catcode`\{=1 \catcode`\}=2 \doi@aux}
\providecommand{\doi@aux}[1]{\endgroup\texttt{#1}}
\makeatother
\providecommand*\mcitethebibliography{\thebibliography}
\csname @ifundefined\endcsname{endmcitethebibliography}
  {\let\endmcitethebibliography\endthebibliography}{}